 \documentclass[apj]{emulateapj}

\usepackage{natbib}
\usepackage{graphicx}
\bibpunct{(}{)}{;}{a}{}{,}% magic!
\newcommand\kmeans{{\em k-means}}
\newcommand\class[1]{ASK\,\,#1}
\newcommand\ask{ASK\,classification}

\shorttitle{Classification
of SDSS/DR7 galaxy spectra}
\shortauthors{S\'anchez~Almeida et al.}

\begin{document}

%% LaTeX will automatically break titles if they run longer than
%% one line. However, you may use \\ to force a line break if
%% you desire.

\title{
Automatic unsupervised classification
of all SDSS/DR7 galaxy spectra
}

%% Use \author, \affil, and the \and command to format
%% author and affiliation information.
%% Note that \email has replaced the old \authoremail command
%% from AASTeX v4.0. You can use \email to mark an email address
%% anywhere in the paper, not just in the front matter.
%% As in the title, use \\ to force line breaks.

%
\author{J.~S\'anchez~Almeida\altaffilmark{1,2},
	J.~A.~L.~Aguerri\altaffilmark{1,2},  
	C.~Mu\~noz-Tu\~n\'on\altaffilmark{1,2},
	and 
	A.~de~Vicente\altaffilmark{1,2}
        }

\altaffiltext{1}{Instituto de Astrof\'\i sica de Canarias, E-38205 La Laguna, Tenerife, Spain
}
\altaffiltext{2}{Departamento de Astrof\'\i sica, Universidad de La Laguna, E-38071 La Laguna,
Tenerife, Spain}
\email{jos@iac.es, jalfonso@iac.es, cmt@iac.es, angelv@iac.es}

%% Mark off your abstract in the ``abstract'' environment. In the manuscript
%% style, abstract will output a Received/Accepted line after the
%% title and affiliation information. No date will appear since the author
%% does not have this information. The dates will be filled in by the
%% editorial office after submission.

\begin{abstract}
Using the \kmeans\ cluster analysis algorithm,
we carry out an unsupervised classification of all 
galaxy spectra in the seventh and final Sloan Digital 
Sky Survey data release (SDSS/DR7). 
Except for 
the shift to restframe 
wavelengths, and the normalization 
to the $g$-band flux,  
no manipulation is applied 
to the original spectra.
The algorithm guarantees that galaxies
with similar spectra belong to the same class.
We find that 99\% of the galaxies can be 
assigned to only 17 major classes, with 
11 additional minor classes including the
remaining 1\%.
The classification is not unique since 
many galaxies appear in between classes, however,
our rendering of the algorithm overcomes this 
weakness with a tool to identify borderline galaxies.
Each class is characterized by a template spectrum, 
which is the average of all the spectra of the galaxies 
in the class. These low noise template spectra vary 
smoothly and continuously along a sequence 
labeled from 0 to 27, from the reddest class to the 
bluest class.
Our  Automatic Spectroscopic K-means-based (ASK) 
classification separates galaxies in colors, 
with classes characteristic of the red sequence,  
the blue cloud, as well as the green valley.
When red sequence galaxies and green valley galaxies 
present emission lines, they are characteristic of 
AGN activity. Blue galaxy classes
have emission lines corresponding to star formation
regions.
We find the expected correlation between spectroscopic class
and Hubble type, but this relationship exhibits a
high intrinsic scatter.
Several potential uses of the ASK classification 
are identified and sketched, including 
fast determination of physical properties  
by interpolation, classes as templates in 
redshift determinations, and target selection 
in follow-up works (we find classes 
of Seyfert galaxies, green valley galaxies, 
as well as a significant number of outliers).
The ASK classification is publicly accessible through 
various websites.
\end{abstract}

%% Keywords should appear after the \end{abstract} command. The uncommented
%% example has been keyed in ApJ style. See the instructions to authors
%% for the journal to which you are submitting your paper to determine
%% what keyword punctuation is appropriate.


%% From the front matter, we move on to the body of the paper.
%% In the first two sections, notice the use of the natbib \citep
%% and \citet commands to identify citations.  The citations are
%% tied to the reference list via symbolic KEYs. The KEY corresponds
%% to the KEY in the \bibitem in the reference list below. We have
%% chosen the first three characters of the first author's name plus
%% the last two numeral of the year of publication as our KEY for
%% each reference.

%% Authors who wish to have the most important objects in their paper
%% linked in the electronic edition to a data center may do so by tagging
%% their objects with \objectname{} or \object{}.  Each macro takes the
%% object name as its required argument. The optional, square-bracket 
%% argument should be used in cases where the data center identification
%% differs from what is to be printed in the paper.  The text appearing 
%% in curly braces is what will appear in print in the published paper. 
%% If the object name is recognized by the data centers, it will be linked
%% in the electronic edition to the object data available at the data centers  
%%
%% Note that for sources with brackets in their names, e.g. [WEG2004] 14h-090,
%% the brackets must be escaped with backslashes when used in the first
%% square-bracket argument, for instance, \object[\[WEG2004\] 14h-090]{90}).
%%  Otherwise, LaTeX will issue an error. 
%%%%%%%%%%%%%
%
\section{Introduction}\label{intro}

{\em The nebulae\footnote{{\rm The galaxies.}} 
are so numerous that they cannot be studied individually. 
Therefore, it is necessary to know whether a fair sample 
can be assembled from the most conspicuous objects and, if so,
the size of the sample required}
\citep[][ Chapter II]{hub36}. Even though
these arguments are from the outset of extragalactic astronomy,
and they refer to the morphological classification of galaxies, 
the reasons put forward by Hubble remain valid today. 
The need to sort out and simplify justify all recent 
efforts to classify the spectra of galaxies 
(\S~\ref{intro.intro}), including the present work.
Such attempts are now more significant 
than ever since we have never had 
the large catalogs of galaxy spectra
available today.
 
The seventh and final Sloan Digital Sky Survey data
release (SDSS/DR7) provides spectra of some 930000 
galaxies \citep[][ and also the SDSS Web 
site\footnote{\tt http://www.sdss.org/dr7}]{sto02,aba09}. 
This 
uniform data set offers a unique opportunity 
to comprehensively classify the different spectra 
existing among nearby galaxies. Our paper presents 
the results of an unsupervised spectral classification 
of all the catalog. Unsupervised implies that 
the algorithm does not have to be trained. It is autonomous 
and self-contained, with minimal subjective influence. 
Thus, we deliberately 
avoid the use of physical constraints, or other 
a priori knowledge. We classify all galaxies
simultaneously, requesting that galaxies with 
similar rest-frame
spectra belong to the same class. 
This approach is in the vein of the rules for a 
good classification discussed by
\citet{san05}, where he points out that physics
must not drive a classification. 
Otherwise the arguments become circular when the 
classification is used to drive physics.
The \kmeans\ algorithm that we implement is 
commonly employed in data mining, machine learning, 
and artificial intelligence \citep[e.g.,][]{eve95,bis06}, but
it has been seldom applied in astronomy 
\citep[see, however,][]{san00}.
From the point of view of the algorithm, 
the galaxy spectra are vectors in a high-dimensional space, 
where they are  distributed among a number of cluster centers.
Each vector is assigned to the cluster whose center is 
nearest, and the center is the average of all the points 
in the  cluster. It works iteratively. Starting from 
guess cluster centers, the spectra are assigned to their 
nearest centers, and then the centers are re-computed until 
convergence is reached. (Further details are given 
in \S~\ref{algorithm}.) 
We choose it because of its extreme computational 
simplicity, as required to deal with large data
sets (\S~\ref{algorithm}), and because
it turned out to work very well in the first case we 
attempted. To our surprise, the algorithm
managed to separate spectra of 
galaxies in the {\em green valley}
within a collection of dwarf galaxies 
encompassing the full range of spectral 
types \citep[][\S~3.1]{san09}. Therefore, we
found it natural to test the ability of  
\kmeans\ to distinguish among all kinds of 
galaxy spectra, and the success of
this follow-up exercise is precisely the work 
reported here. In addition to the above virtues, the 
\kmeans\ method provides a prototypical high 
signal-to-noise spectrum for each class of galaxy,
being the spectra of the galaxies in a class
similar to the associated prototypical spectrum.
These few representative spectra can be studied and 
characterized in detail as if they were individual galaxies, 
and then their properties can be attributed to all class 
members (\S~\ref{applications}). Other popular classification 
methods lack this powerful and convenient 
feature (see \S~\ref{intro.intro}).

The acronym ASK stands for 
Automatic Spectroscopic K-means-based, 
and it is used throughout the text to denote our 
classification. The paper is structured as follows.
\S~\ref{intro.intro} provides an overview of the main 
spectral classification methods employed so far.
It also summarizes systematic trends resulting 
from the application of those methods. 
Our \kmeans\ classification algorithm is examined in
\S~\ref{algorithm}, where we test the class recovery
upon known classes (\S~\ref{test_qbcd}), we analyze 
the repeatability of the classification (\S~\ref{recovery}), and
we assign probabilities to class membership 
(\S~\ref{assigning}). 
The SDSS/DR7 dataset is briefly introduced in 
\S~\ref{data_set}. The actual  classification of  
SDSS/DR7 is described in \S~\ref{final_imp}.
The \ask\  is compared with Principal Component 
Analysis (PCA) classification in \S~\ref{pca_class} 
(see also \S~\ref{intro.intro}).
The self-consistency of the \ask\
is discussed in various 
sections dealing with specific results;
relationship between ASK class and Hubble type (\S~\ref{ask_class_vs_morph}), 
ASK class and color sequence (\S~\ref{colorscolors}), 
ASK class and AGN activity (\S~\ref{ask_vs_agn}), 
and ASK class and redshift (\S~\ref{ask_vs_redshift}).
Further applications of the classification procedure
are sketched in \S~\ref{applications}.
The \ask\ is publicly available as we explain 
in \S~\ref{conclusions}. This section also outlines 
ongoing works based on ASK.

\subsection{Spectral classification of galaxies}\label{intro.intro}

The first spectral classifications of galaxies are almost
coeval with the discovery of the Hubble sequence.
\citet{hub36} discusses how the spectral types and 
colors systematically vary within the morphological sequence, 
being ellipticals the reddest
and open spirals the bluest \citep[see also][]{hum31}.
One of the early attempts to set up 
an spectroscopic classification of galaxies is that by 
\citet{mor57}. They assign the blue part of the visible 
spectrum  (3850\,\AA\ -- 4100\,\AA ) to stellar classes 
from A to K. They find a clear relationship 
between spectral class and shape, 
with the most concentrated galaxies (E, S0) belonging
to class K, and the most diffuse galaxies (Sc, Irr)
included in class A. The relationship applies to some 80\% of the
galaxies, a percentage probably larger for the targets of
highest luminosity.
\citet{aar78} shows how the visible and IR colors of galaxies
along the Hubble sequence can be understood as a one parameter
family, in terms of the superposition of spectra of
A0V dwarf stars and M0III gigant stars.
\citet{ber95} points out that a simple model consisting
of two stellar spectral types can reproduce the observed
broad band colors, but only if the spectral types are
allowed to vary. Five primary spectral types 
result from this modeling.
Similar conclusions are also reached by \citet{zar95}
using stellar spectrum fitting.

Principal Component Analysis (PCA) is probably the most
popular classification method employed so far.
Each spectrum is decomposed as a linear
superposition of a small number of eigenspectra, so that
a few coefficients  in this expansion (eigenvalues) 
fully describe the spectrum. It is fairly fast and robust, and 
a solid mathematical theory supports it \citep[][]{eve95}. 
To the best of our knowledge, the first applications of 
PCA in this field have to do with stellar classification
\citep[e.g.,][]{dee64,whi83}, then moved to
quasar spectra \citep[e.g.,][]{mit90,fra92}, and finally 
arrived to the spectral classification of regular
galaxies \citep[e.g.,][]{sod94,con95}. 
PCA is the method  of reference, and  
we compare it in \S~\ref{pca_class} with our \kmeans .
Two general results are common to all PCA analyses.
Spectrumwise, galaxies can be characterized 
and distinguished by means of a single parameter
that links the coefficients of the two or three 
first eigenspectra. Then different classes
are obtained by splitting 
(somewhat artificially)
this 1-dimensional  family into pieces.
The approach holds for 2dF galaxies 
\citep{fol99,mad03},
for galaxies in \citet{ken92} \citep{con95,sod97},
for Las Campanas Redshift Survey galaxies \citep{bro98},
for DEEP2 galaxies \citep{mad03b}, 
for IUE galaxies \citep{for04}, 
and for SDSS \citep{yip04}.
The second common result is the 
correspondence between
spectral sequence and Hubble type. 
Even though ellipticals tend to be 
red and spirals tend to be blue,
such relationship has a large 
intrinsic scatter \citep{con95,sod97,fer06b},
which augments towards the UV \citep{for04}. 
Sometimes elliptical galaxies with
blue colors are found in the local 
universe \citep[e.g.,][]{kan09},
and this deviation from the trend 
is expected
to grow even further with increasing 
redshift if, as \citet{con06} 
argues, it is a coincidence that Hubble types
correlate with colour in the nearby universe. 
At higher redshifts morphologically classified 
ellipticals are often blue in colour and actively 
forming stars \citep{con06,hue09}.

Despite the advantages mentioned above, 
PCA presents a clear drawback. It does not provide
prototypical spectra to characterize the classes.
The  PCA eigenspectra do not resemble 
any member of the data to be classified and, 
in general, eigenspectra are of difficult
physical interpretation
\citep[e.g.,][]{cha03,for04,yip04}. 
The advantage of having classes characterized
by prototypical spectra is clear. 
These few spectra can be studied in detail
using standard diagnostic techniques developed for 
individual galaxies through the years. 
Then the attributes of the prototypical spectra
can be passed on to the class members, or 
they can be used as intermediate grid-points
to interpolate the properties of the 
class members (see \S~\ref{applications}). Moreover,
the differences between a particular 
galaxy and its class prototype 
allow for precise relative measurements. 
In an attempt to complement PCA with this feature,
\citet{cha03} developed an {\em archetypal analysis}
algorithm.  As the authors explain, it is like PCA but
the eigenspectra are required to be members or 
mixtures of members of the input data set. However,
by construction, the eigenspectra are extreme data 
points lying on the data set outskirts. 
Although physically meaningful, the 
eigenspectra  are outliers, and it may be difficult
to connect their physical properties with those of 
typical galaxy spectra. Other improvements on the basic
PCA technique are local linear embedding 
\citep{van09}, and ensemble learning independent
component analysis \citep{lu06}.
These extensions are  computationally expensive, 
and so far they have been introduced only as
proof-of-concept works. 

In addition to the superposition of stellar spectra
and the PCA techniques described above, 
galaxies have been classified using 
neuronal networks 
\citep{fol96,mad03}, massive lostless data compression
\citep{rei01}, information bottleneck \citep{slo01},
and probably others. 
The algorithms have flourished in response
to the availability of  new large spectral 
databases. We are still in an expanding phase, which
should lead to a final convergence of the various techniques.
The different methods seem to roughly coincide 
in the global picture, but it is so far unclear whether
they agree in the details.

\section{The classification algorithm}\label{algorithm}

In the context of classification
algorithms, galaxy spectra are vectors in 
a high dimensional space, with as many dimensions as
the number of wavelengths in use. The galaxy catalog 
to be classified is a set of vectors in this space,
and so the (Euclidean) distance between
any pair of them is well defined. Vectors (i.e.,
spectra) are assumed to be clustered around 
a number of cluster centers. The classification
problem consists in (a) finding the number
of clusters,
(b) finding the cluster centers, 
and (c) assigning each galaxy in the catalog to one 
of these centers. 
We employ the \kmeans\ algorithm to carry out this
classification 
\citep[see, e.g., Chapter~5 in ][]{eve95,bra98,san09}.
In the standard formulation, it begins by selecting at 
random from the full data set a number $k$ of 
template spectra. Each template spectrum 
is assumed to be the center of a cluster, and each 
spectrum of the data  set is assigned to the 
closest cluster center (i.e., that of minimum
distance or, equivalently,
closest in a least squares sense). 
Once all spectra in the dataset
have been classified, the cluster center is re-computed 
as the average of the spectra in the cluster. This
procedure is iterated with the new cluster centers, 
and it finishes when no spectrum is re-classified 
in two consecutive steps. 
The number of clusters $k$ is arbitrarily chosen but,
in practice, the results are insensitive to such selection
since only a few clusters possess 
a significant number of
members, so that the rest can be discarded.
On exit, the algorithm provides a number of 
clusters, their corresponding cluster 
centers, as well as the classification 
of all the original spectra now 
assigned to one of the clusters. 

The algorithm is simple and fast, as
required to treat large data sets. It
assures that galaxies with similar 
spectra end up in the same cluster,
and provides cluster centers, i.e.,
prototypical spectra representative
of all the galaxies in a cluster. In
addition, it seems to work very well separating
galaxy spectra, as inferred from the  first test 
(see \S~\ref{intro}), and from this work.
Unfortunately, it has a major drawback.
It yields different clusters with each random 
initialization. After pondering pros and cons, 
we decided to carry on with the algorithm, 
but not without evaluating the impact of 
the initialization on the classification.
The impact is quantified and controlled through
three complementary methods: 
(1) carrying out different random initializations
and comparing their results,
(2) assigning galaxies to several 
classes, each one with its own probability,
and (3) trying alternative methods of 
initialization. The first point is dealt with 
in \S~\ref{test_qbcd} and \S~\ref{final_imp},
leading to classifications whose
classes share some 70\% of the galaxies. 
It is not 100\% because of spectra lying
in between classes. This difficulty is to some extent
cured by the second point, treated in 
\S~\ref{assigning}, which allows us to
assign galaxies to several classes and,
therefore, to 
identify galaxies in class borders.
The third point is treated in the next 
paragraph, concluding that the scatter in the 
classification is not significantly 
modified by the mode of initialization. 
It does modify the timing, though.

We tried several initialization methods, including the 
standard one, that by \citet{bra98}, and others \citep{pen99}.
None of them seem to reduce the scatter
due to the initial random seed (\S~\ref{test_qbcd} 
and \S~\ref{final_imp}). This behavior can be 
understood in terms of the existence of
borderline galaxies, as argued 
in Appendix~\ref{change_center}.
A proper initialization, however, reduces the
iterations required to converge, and speeds up the 
procedure. We have adopted our own method, which is 
simple and fast, and it starts off with a reduced
number of classes. It tries to select initial cluster 
centers according to the clusters that exist in the
data set. If initial centers are purely chosen at random, 
then the clusters having the largest number of 
elements are overrepresented, and minor clusters 
may be even absent. The procedure works as follows:
(Step 1) choose at random a small set of  initial cluster 
centers (say, 10). (Step 2) Run one iteration 
of the standard \kmeans , 
and select as initial cluster center 
the cluster center with  the largest number 
of elements. (Step 3) Remove from the 
set of galaxies to be classified  those 
belonging to the selected cluster center. 
(Step 4) Go to step 1 if galaxies are still left;
otherwise end.
In addition to this particular initialization, 
we tuned the standard \kmeans\ described above 
with one extra ingredient. The iteration loop ends 
when the classifications in two successive steps are 
sufficiently close one another, i.e., when 99\% of the 
assignations do not vary between two iterations. 
This simplification
speeds up the convergence since the classification of 
the remaining 1\% takes a long time, and
does not help finding the main galaxy 
classes, already well characterized by 99\% of the 
sample.

\subsection{Testing the repeatability 
of the classification}
\label{test_qbcd}

Two classifications of the same data set are
identical if they include the same galaxies
in each class. With this criterion in mind, we 
compare two different classifications by 
pairing their two sets of classes according to the 
number of galaxies that they have in common. 
We compute the number of galaxies in common
between each pair of classes formed by  
one class from one classification and 
the second class from the second classification.
The two classes sharing the largest 
number of galaxies are assumed to be {\em equivalent}.
The same criterion is repeated until all the classes 
of one of the classifications have been paired. 
Since the number of classes in the two classifications 
are not necessarily the same, some classes remain
unpaired. This procedure tries to maximize the
% guarantees a maximum 
number of galaxies sharing
the same class in the two classifications.
We  use the percentage of galaxies in
equivalent classes as a measurement of the
agreement between the two classification, dubbing
it coincidence rate.

A first series of tests to check repeatability
has been carried out using the 21493 
quiescent blue compact dwarf galaxies (QBCD)
selected by \citet{san09} from SDSS/DR6. 
Here we employ the same wavelength 
windows used in
\citeauthor{san09}~(\citeyear{san09}; they are labeled as
QBCD in Table~\ref{tab1}). 
Thirty independent classifications of this 
data set yield a coincidence rate of 
$71\pm 9$~\%, with the error bar being the 
standard deviation. We think that
the origin of these fluctuations in the number of 
common galaxies is due to the random 
initialization coupled with the large number
of variables defining the spectra; 
see Appendix~\ref{change_center}.
This 70\% coincidence has two consequences. 
(1) A galaxy chosen at random from the sample
has a 70\% chance of appearing in 
equivalent classes
in two different runs of the classification.
(2) The cluster centers are very well defined since, 
independently of the initialization, they share 70\%
of the galaxies that define them. The coincidence of 
the final classification of SDSS/DR7 is similar 
(but a bit smaller), as it is discussed in detail
in \S~\ref{final_imp}.

\subsection{Testing the class recovery upon known classes}
\label{recovery}

We construct a set of mock observations to see whether the
algorithm is able to recover clusters imposed on the data.
To a data set of 21493 spectra (same number
as in \S~\ref{test_qbcd}), we add different amounts of 
pixel-to-pixel uncorrelated random noise. The
21493 spectra were randomly selected among
three real spectra representative of galaxies
along the color sequence 
in the blue  cloud, the red sequence, and the green valley
(see \S~\ref{colorscolors}). 
This 3-class mock observation encompasses 
the full range of spectra to be expected.  
When no noise is added, 
the algorithm returns 3 classes, with
100~\% coincidence between the original and 
the classified spectra.
As the noise increases, the number of recovered classes
increases too. The fact that new classes have appeared does not 
mean that the algorithm is malfunctioning. 
Actually, the algorithm seems to be classifying the noise. 
When the signal-to-noise ratio per pixel is $\simeq\,$10, 
which is typical of SDSS spectra, one retrieves some 10 
classes.  However, the spectra of the different 
original classes are never mixed up, i.e., they end up in 
separate classes. The noise artificially increases 
the number of classes,  but it does not wash them out.
Moreover, it is easy to figure out which classes are
faked  by this kind of random noise because, 
being pixel-to-pixel uncorrelated,
it does not modify global 
properties of the spectra such as colors. 
Different classes with different colors are not 
artifacts created by noise.

\subsection{Assigning several classes to each galaxy}
\label{assigning}

Given a collection of galaxy spectra, the \kmeans\
algorithm infers a small set of classes or clusters,
and assigns each spectrum to one of them. 
A number of reasons advice addressing 
the inverse problem too, i.e., assigning classes to 
individual spectra once the classes are known.
This alternative is required to classify spectra 
not used  in the classification, which turns out to
be a practical case of major interest 
(e.g., \S~\ref{final_imp}, \S~\ref{ask_class_vs_morph}).
In addition, the classification does not provide 
unique sharp classes. The results in \S~\ref{test_qbcd} 
suggest that a significant number of galaxies 
are in between classes, and this fact
could be easily acknowledged and quantified  
with a procedure to estimate the probability that 
a given galaxy belongs to each one of 
the known classes. Borderline galaxies must fit in 
several classes with similar probabilities. 
A general procedure to carry out such multiple
assignation is worked out in this section.

The distance of spectrum  ${\bf s}=(s_1,s_2,\dots)$
to class center ${\bf c}=(c_1,c_2,\dots)$ is defined as,
\begin{equation}
d({\bf s}-{\bf c})=|{\bf s}-{\bf c}|=\big[\sum_iw_i\,(s_i-c_i)^2\big]^{1/2},
\label{def_dispersion}
\end{equation}
where $s_i$ and $c_i$ are the values of the 
spectrum and the cluster center in the $i-$th 
wavelength pixel.
The weights $w_i$ allow one to select a subset
among the pixels defining the spectra, i.e.,
\begin{equation}
w_i=\cases{0, & in discarded wavelengths,\cr
	m^{-1},& in used wavelengths,}
\end{equation}
with $m$ the total number of pixels where $w_i\not= 0$.
\kmeans\ selects as class center 
the average of all the spectra belonging
to the class. Each one of these spectra 
has its own distance to the cluster center, so that
the full set defines 
a distribution of distances to the cluster center for the
spectra in the class. Let us call $f_{\bf c}(d)$ 
the probability density function (PDF) of such distribution, 
i.e., $f_{\bf c}(d)\,\Delta d$ is
the probability of finding a galaxy
in cluster {\bf c} with a distance to the  
cluster center between $d$ and $d+\Delta d$.
The chances that galaxy {\bf s}
belongs to cluster {\bf c} can be estimated 
as the probability of finding galaxies in the
cluster with
distances equal or larger than 
$d({\bf s}-{\bf c})$, i.e., 
\begin{equation}
P({\bf s},{\bf c})=
\int_{d({\bf s - c})}^\infty f_{\bf c}(x)\, dx.
\end{equation}
Unfortunately, sorting the classes 
according to their $P({\bf s},{\bf c})$ may be 
inconsistent 
with the assignation of classes made by \kmeans . 
Given a galaxy spectrum, \kmeans\ assigns to it 
the class of minimum distance,  i.e.,
the class ${\bf c}_k$ where 
$d({\bf s}-{\bf c}_k) \leq 
d({\bf s - c})\, \forall\, {\bf c}$. However, 
there is no guarantee that the class of minimum 
distance coincides with the class of maximum probability, 
i.e., in general 
${\bf c}_k\not={\bf c}_p$ with ${\bf c}_p$ defined
so that
$P({\bf s}, {\bf c}_p) \geq 
P({\bf s}, {\bf c})\,
\forall\, {\bf c}$.
Consequently, the sorting of classes according to
their probabilities cannot be used
to order classes in a way that agrees with \kmeans .
We circumvent the problem defining a merit function,
based on probabilities, that can be used to judge 
the membership of a given galaxy to the 
various classes. As a main constraint,
the class of minimum distance must have
the largest merit, so that the ordering provided
by this merit function agrees with the assignation
made by the \kmeans\ algorithm.

We call such merit function {\em quality}.  
For a spectrum {\bf s}, assigned by \kmeans\ 
to class ${\bf c}_k$, the quality of
class ${\bf c}$ is defined as,
 \begin{equation}
Q({\bf c},{\bf s},{\bf c}_k)=\int_{d({\bf s- c})}^\infty f_{{\bf c}_k}(x)\, dx.
\label{prop1}
\end{equation}
Given a spectrum, its membership to various classes is 
judged according to $Q$, with the class of largest
$Q$ the main affiliation, the one of second largest $Q$,
the second main, and so on.
The sorting according to quality is actually a sorting
according to distance to the cluster centers since, 
\begin{equation}
Q({\bf c_1},{\bf s},{\bf c}_k) \ge Q({\bf c_2},{\bf s},{\bf c}_k)
\Leftrightarrow
d({\bf s - c_1}) \le d({\bf s - c_2}).
        \label{ecu1}
\end{equation}
This property guarantees that the class of minimum distance 
has the largest quality and, therefore,
the main class attending to its quality 
agrees with the \kmeans\ assignation.
Equation~(\ref{ecu1}) follows from Eq.~(\ref{prop1})
because $f_{\bf c}$ is always positive, and so, $Q$ 
is a monotonic decreasing function of $d({\bf s -c})$.
In addition to conforming to \kmeans ,
the quality $Q$ has a number of 
practical properties.
Due to the normalization of the PDFs,
\begin{equation}
\int_0^\infty f_{\bf c}(x)\,dx=1,
\end{equation}
the quality is always comprised between zero, 
for no match, and one, for perfect match,
\begin{equation}
Q({\bf c},{\bf s},{\bf c}_k)\rightarrow\cases{
         1, & when $d({\bf c - s})\rightarrow 0$,\cr
        0, & when $d({\bf c - s})\rightarrow \infty$.
        }
\end{equation}
The best quality, i.e., that of the class of minimum 
distance, admits a simple interpretation. It
is just the probability that the galaxy belongs 
to this class because,
\begin{equation}
Q({\bf c}_k,{\bf s},{\bf c}_k)=P({\bf s},{\bf c}_k).
\end{equation}
If the best quality is large, then there
are high chances that the galaxy is part of the 
class. If the best quality is similar to the second
best quality, then the galaxy is in between classes.
If the best quality is very small ($\ll 1$), then 
the galaxy is an outlier, meaning that it
does not fit into any of the main classes. 

\begin{figure}
\includegraphics[width=0.45\textwidth]{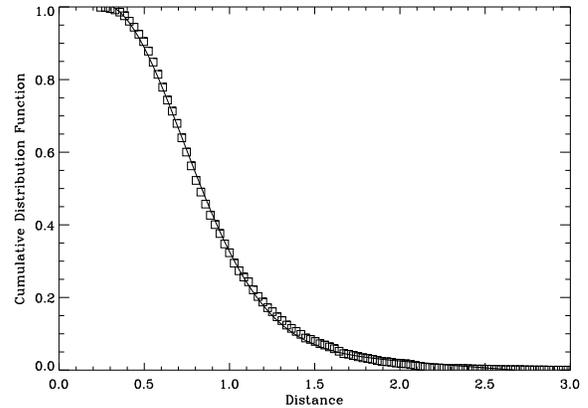}
\caption{Example of cumulative distribution function of distances
to the class center. The symbols show the observed values 
whereas the solid line corresponds to the 
analytical representation used in the work.
Distances are relative to the standard 
deviation.}
\label{cumula}
\end{figure}

Computing qualities as explained above requires estimating
the PDF for the distribution of distances within a class.
This can be derived from the histogram of distances
to the cluster center among the galaxies that \kmeans\ 
has included within the class.
After several trials, 
we found out that the observed
cumulative distribution of distances can be  
very well approximated as, 
\begin{equation}
\int_{d}^\infty f_{\bf c}(x)\, dx\simeq a_1+a_2\,d+
        a_3\,\exp\big[-[(d-a_4)/a_5]^2\big],
\end{equation}
where the five free coefficients 
$(a_1,\dots, a_5)$ are determined by a non-linear 
least-squares fit of the empirical cumulative
histogram of observed distances.
Figure~\ref{cumula} shows one of such fits.

\begin{figure}
\includegraphics[width=0.5\textwidth]{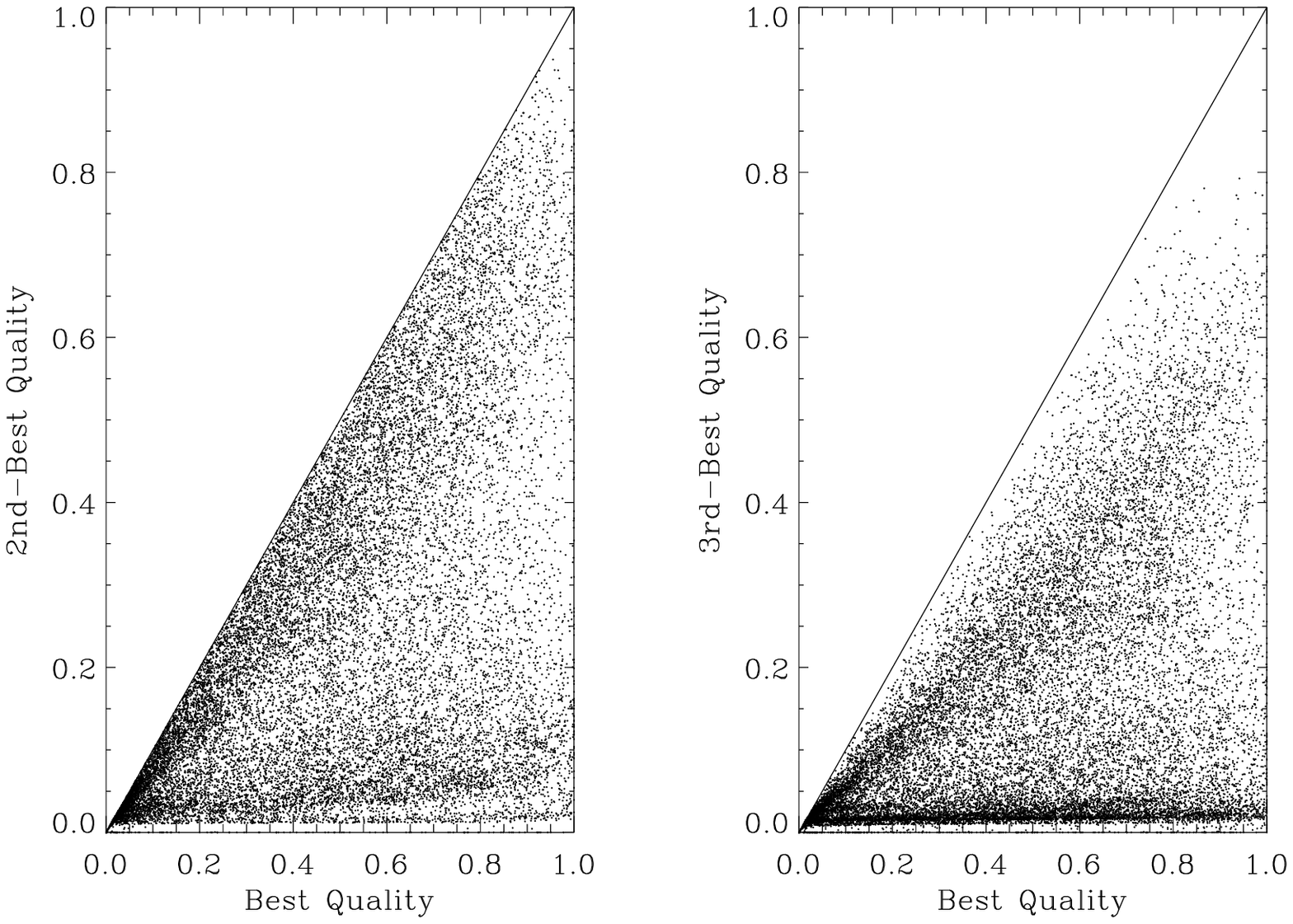}
\includegraphics[width=0.5\textwidth]{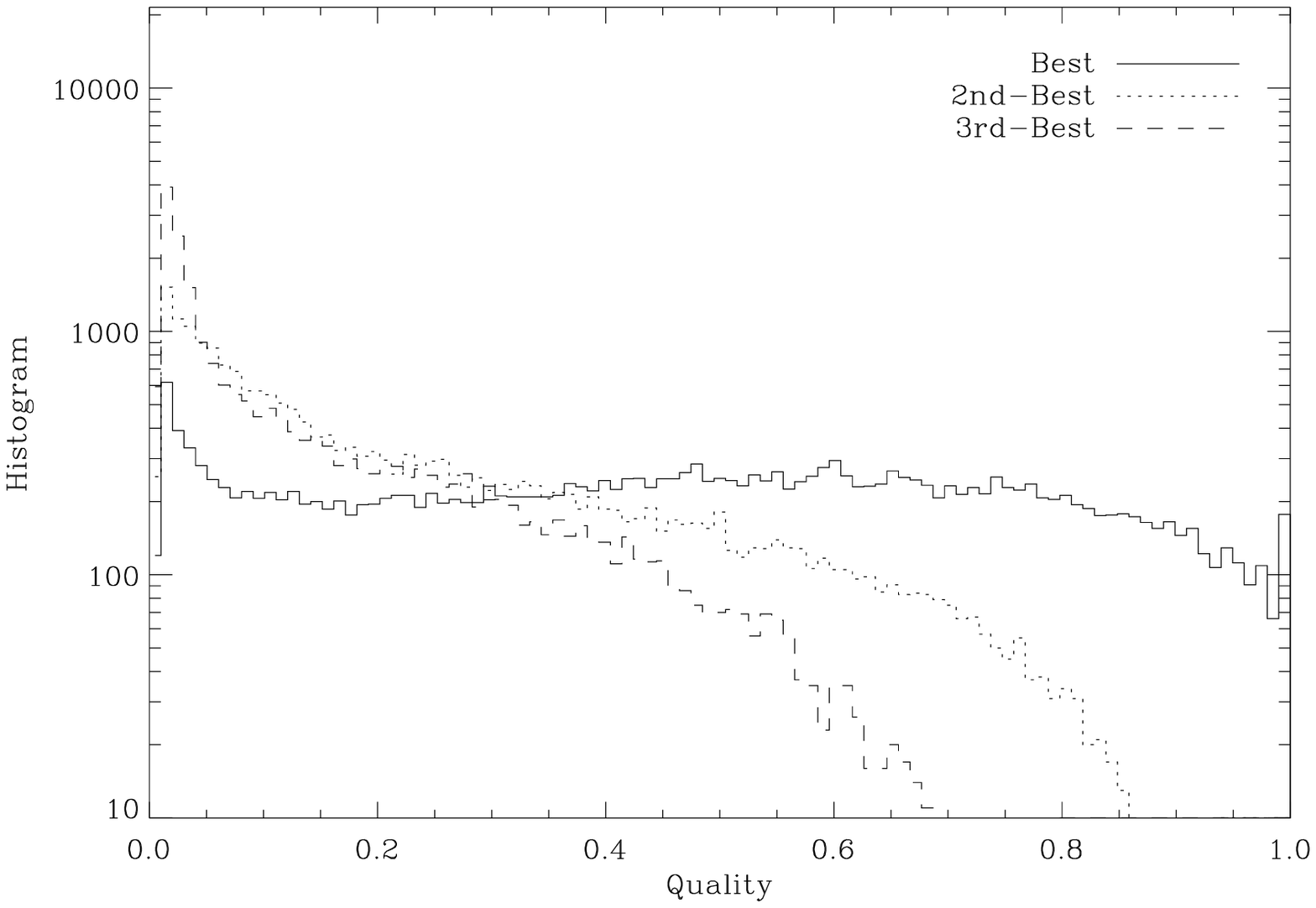}
\caption{Top: example of scatter plot of quality versus quality.
Top-left: 2nd-best quality versus best quality. Points not far
from the diagonal can be classified as either
one of the two classes. 
Top-right: 3rd-best quality versus best quality.
Bottom: histograms of qualities for the best quality 
class (the solid line), the 2nd-best class (the dotted line), 
and the 3rd best
class (the dashed line). Note that even the best
qualities are sometimes close to zero, implying that
these galaxies are outliers of the classification.
}
\label{set_quality1}
\end{figure}

Real qualities are represented in Fig.~\ref{set_quality1},
which includes the results in one of the test 
classifications in \S~\ref{test_qbcd}.
Figure~\ref{set_quality1}, top, shows scatter
plots of the 2nd-best quality versus the best quality 
(top left), and the 3rd-best quality
versus the best quality (top right). 
A significant number of galaxies appear not far
from the border where the best and 2nd-best 
qualities are equal and, therefore, many
galaxies lie in between classes.
Figure~\ref{set_quality1}, bottom, shows 
the histograms of those qualities. 
Note how the best quality has a 
rather flat distribution peaking at, say,
0.6. Note also how the number of galaxies 
with very small best quality is also very 
small. These are outliers whose spectra 
differ significantly from the characteristic 
spectra of the main classes. 
 
\section{The data set: SDSS/DR7 spectra}\label{data_set}

The SDSS/DR7 is the final major data realease 
of the SDSS project. Details about SDSS and the DR7 
can be found in, e.g., 
\citet{sto02}, \citet{aba09}, and also in the 
thorough 
SDSS~website\footnote{{\tt http://www.sdss.org/dr7}}.
The spectroscopic part of the survey 
contains some 930000 galaxy spectra,  
and this full set is classified in our work.
The basic properties of spectrograph and 
spectra will be summarized here, 
but we refer to the references given above for 
further details.
The SDSS spectrograph has two independent 
arms, with a dichroic separating
the blue beam and the red beam at 6150 \AA .
It  simultaneously renders a spectral range 
from 3800\,\AA\ to 9250\,\AA , with 
a spectral resolution between 1800
and 2200. The sampling is linear in 
logarithmic wavelength, with a mean
dispersion of 1.1\,\AA\,pix$^{-1}$ in the blue 
and 1.8\,\AA\,pix$^{-1}$ in the red. 
Repeated 15 min exposure spectra are 
integrated to yield a S/N per pixel $>4$  
when the apparent magnitude in the  
$g$-band is 20.2.
The spectrograph is fed by 
fibers which subtend about 3\arcsec
on the sky. Most galaxies are larger than
this size therefore the fibers tent to 
sample their central regions 
(e.g., 88\% of the galaxies have effective 
radii larger than half the fiber 
diameter).

Two adjustments are made on the
original spectra before classification.
First, they are brought to restframe
wavelengths using the redshifts provided by SDSS.
This wavelength shift involves an interpolation,
and we take advantage of this need to bring to
a common wavelength scale all spectra,
as required by the classification algorithm.
The common scale has the same number of pixels 
as the original spectra (3850), and it
is equispaced in logarithmic wavelength from 
3800\,\AA\ to 9250\,\AA . Obviously, the 
IR part of the spectrum is missing
as the redshift increases, and we 
extrapolate it with a constant. Note, however,
that this missing part is not used for
classification (see below and \S~\ref{final_imp}).
The second manipulation is a 
global scaling
applied after the resframe correction.
The spectra are
normalized to the flux in the $g$~color 
filter (effective wavelength $\simeq$ 4825\,\AA ),
a normalization factor that we compute for
each spectra using the transmission curve
provided by SDSS.
This re-scaling automatically corrects for
the flux dimming associated with the redshift
\citep[e.g.,][]{bla07},
but the original motivation was allowing
comparison between galaxies 
of different absolute magnitudes. If the
global scaling is not removed, the flux of the galaxy 
completely dominates the classification, and 
galaxies are split in bins of equal luminosity,
rather than in spectral classes.

No further correction has been applied to the 
data. We do not correct for extinction, seeing,
galaxy size, aperture bias, etc. This apparent
sloppiness actually results from 
a deliberate attitude towards classification,  
following the guidelines by \citet{san05} mentioned
in \S~\ref{intro}. If these corrections are 
important and  the classification is working properly, 
then the spectra of the same type of galaxy with and
without an uncorrected bias should appear 
in separate bins. It is then a matter of 
a posteriori physical interpretation 
to infer what causes the different classes,
and eventually join some of them when appropriate.

%%%%%%%%%%%%%%%%%%%%%%%%%%%%%%
\section{Final implementation: the classification of SDSS/DR7}
\label{final_imp}

\begin{figure}
\includegraphics[width=0.5\textwidth]{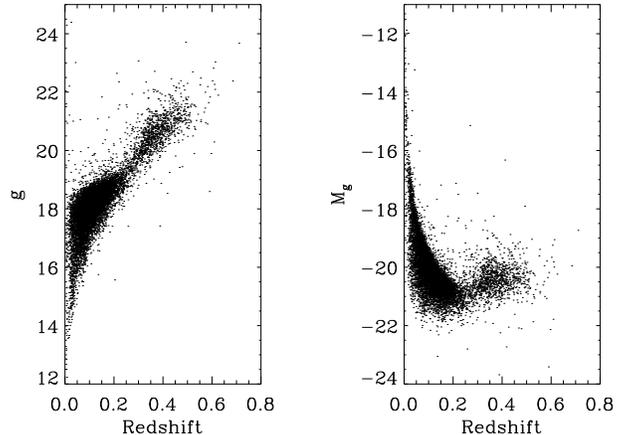}
\caption{Scatter plots of apparent $g$ magnitude (left)
and absolute $g$ magnitude (right) versus redshift for 
the galaxies in the SDSS/RD7. The plots include 
only a fraction of the galaxies chosen at random to
avoid overcrowding.
}
\label{scatter1}
\end{figure}

The spectra to be classified by \kmeans\ 
must share the same wavelength scale, i.e.,
the same sampling interval and the 
same wavelength range. 
SDSS/DR7 has a significant number of galaxies up to
redshift 0.5 (see Fig.~\ref{scatter1}). At this redshift 
the reddest restframe wavelength that
SDSS provides is 6200\,\AA  , therefore, in order
to use the full data set for classification,
one should restrict the range of 
wavelengths down to 6200\,\AA . Alternatively, one 
can restrict the range of redshifts of the galaxies. 
We have chosen the second possibility to avoid  
overlooking in the classification
lines as important as H$\alpha$. 
The full set of
spectra described in \S~\ref{data_set}
has been divided into a low redshift
part (redshift $\leq 0.25$, with 788677 spectra)
and a high redshift part 
(redshift $> 0.25$, with 138649 spectra).
The low redshift part is classified
by means of the \kmeans\ algorithm, 
which provides the classes. 
Then the high redshift 
part is classified according to the classes 
derived from the low redshift part using 
the tools developed in \S~\ref{assigning}.
The reason for choosing 0.25 as the dividing
redshift is twofold. First, the distribution
of redshifts in SDSS/DR7 seems to present a 
discontinuous behavior at roughly
this redshift (see Fig.~\ref{scatter1},
and also the discussion in \S~\ref{ask_vs_redshift}). 
Second, and most important,  0.25 is the 
largest redshift that allows us to include
for classification the near-IR {\sc TiO} bands 
characteristic of M stars  -- the reddest restframe 
wavelength at this redshift is some 7500\,\AA.

\begin{figure*}
\includegraphics[width=.8\textwidth]{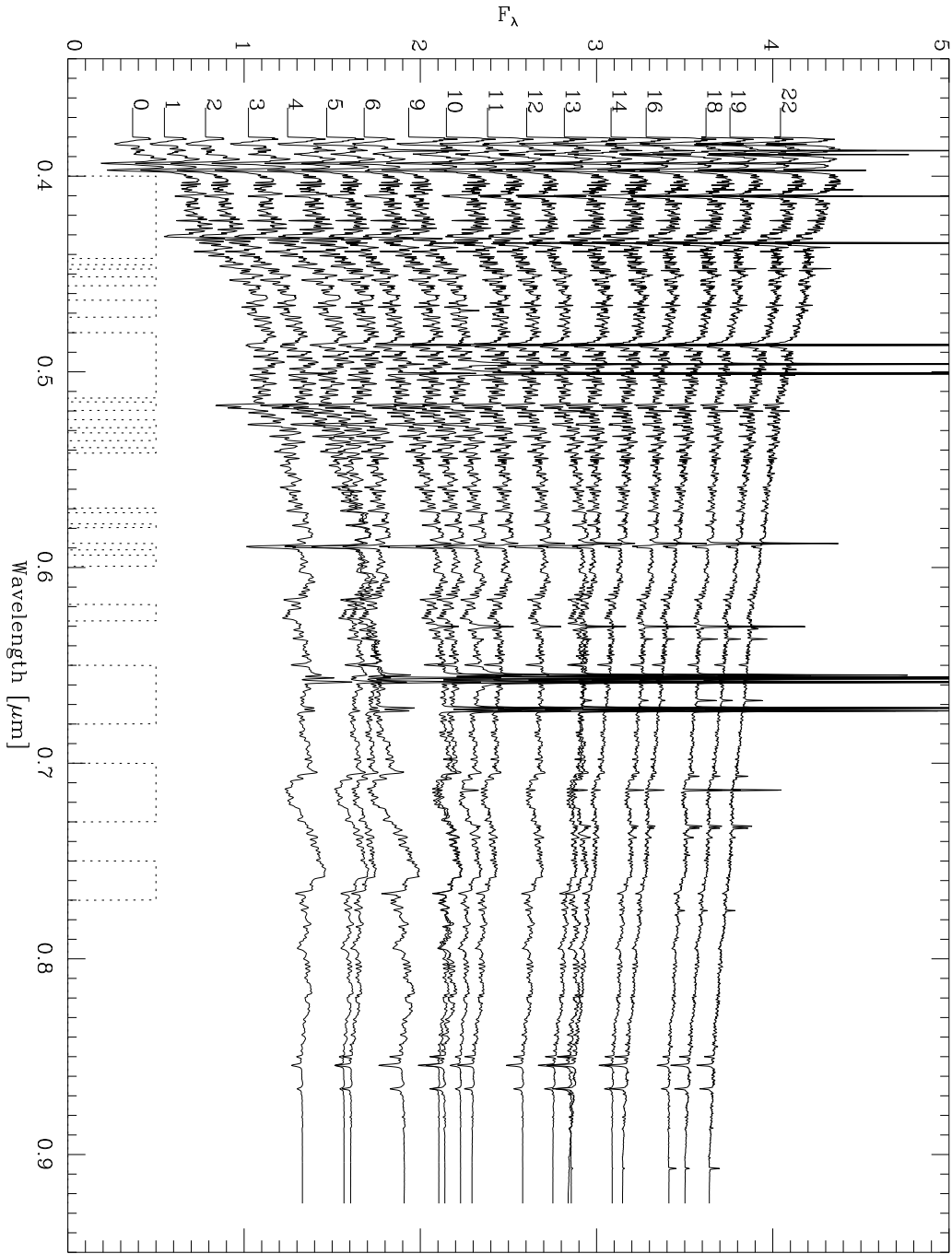} % for referee format 
\caption{
Template spectra representing the 
major classes in the ASK classification of 
the SDSS/DR7 galaxies. The different spectra
have been artificially shifted upward 
according to their $u-g$ color. (Otherwise
the plot becomes overcrowded.)
The numbers next to the spectra (in the left
hand side of the plot) correspond to the
class number, which was assigned according
to the $u-g$ color (\class{0}
for the reddest, \class{1} for the second reddest, and so on
up to \class{22}).
Gaps in the numbering indicate the presence of 
minor classes of intermediate colors.  
The fluxes are in dimensionless units, i.e.,
they are normalized to the average 
flux in the $g$-filter bandpass.
Wavelengths are given in $\mu$m.
The dotted line differs from
zero at the  wavelengths used for 
classification. 
}
\label{classification}
\end{figure*}

In addition to removing the reddest part of the 
spectrum missed by the redshift of the galaxies,
several reasons advice using only selected bandpasses 
to carry out the 
classification. Including too much continuum 
does not add information but dilutes the signals contained 
in the spectral lines. The number of wavelengths in the
spectra sets the dimensions of the vectors to be 
classified. The larger the number of wavelengths the
more computationally  demanding the classification, which
makes it advisable
limiting the number of wavelengths. 
Keeping these caveats in mind, we use
for classification only the bandpasses shown 
as dotted lines  in the bottom of  
Fig.~\ref{classification}, 
which are also listed in Table~\ref{tab1}. 
Except for a near-IR window between 8400\,\AA\ and 8800\,\AA ,
they include all the bandpasses employed by 
\citet[][\S~3]{san09} in the classification
that triggered the present work 
(see \S~\ref{intro}).
These bandpasses contain the main 
emission lines that trace activity 
(star formation and 
AGN activity). Since they are distributed along the 
visible spectrum, they also provide sensitivity 
to the colors of the galaxies.
In addition, we include all the bandpasses of the 
Lick indexes, which were selected because they 
depend on the age and metallicity of the 
stellar content of the galaxies\footnote{Using 
this argument to select 
bandpasses somehow conflicts with the philosophy 
of having a classification not driven by physics.
However the conflict is only marginal. 
The lick indexes cover a large part of the 
spectrum, and we take all of them blindly.
Using the Lick bandpasses is only a  particular 
way of enhancing the contribution of spectral 
lines with respect to continuum.}\citep{wor94,wor97}.
Finally, we include two windows at the location
of {\sc TiO} bands characteristic of M stars 
and early type galaxies (at 7150\,\AA\
and 7600\,\AA ).
These bandpasses are sensitive to
the level of the near-IR continuum, and 
are tracers of old stellar populations.

%% new table as suggested by alfonso
%
% prepared using create_table_weight.pro
%
\begin{deluxetable}{cl}
\tablecolumns{2}
\tablewidth{0pc}
\tablecaption{Bandpasses used in the ASK 
classification of SDSS/DR7}
\tablehead{
\colhead{From -- To}    &  
\colhead{Comment}
}
\startdata
4000 -- 4420 & QBCD blue, H$\delta_A$, H$\delta_F$,
        CN$_1$, CN$_2$,\\
        & Ca4227, G4300, H$\gamma_A$, H$\gamma_F$, Fe4383 \\
4452 -- 4474 &               Ca4455 \\
4514 -- 4559 &               Fe4531 \\
4634 -- 4720 &               Fe4668 \\
4800 -- 5134 &  QBCD green, H$\beta$, Fe5015, Mg$_1$ \\
5154 -- 5196 &                 Mg$_2$, Mg$_b$ \\
5245 -- 5285 &               Fe5270 \\
5312 -- 5352 &               Fe5335 \\
5387 -- 5415 &               Fe5406 \\
5696 -- 5720 &               Fe5709 \\
5776 -- 5796 &               Fe5782 \\
5876 -- 5909 &                 Na~D \\
5936 -- 5994 &                TiO$_1$ \\
6189 -- 6272 &                TiO$_2$ \\
6500 -- 6800 &             QBCD red \\
7000 -- 7300 &             TiO band \\
7500 -- 7700 &             TiO band
\enddata
\tablecomments{Wavelengths are in \AA. The {\em Comment} contains
the names of the Lick indexes in the bandpass,
plus additional information used to identify 
the bands in the main text.}
\label{tab1}
\end{deluxetable}

Initially, 
the computer resources needed to carry out the 
classification were unclear. The procedure 
is iterative (\S~\ref{algorithm}), and the timing 
is mostly set by the number of iterations, 
which scales in a unknown fashion with the number 
of spectra and wavelengths ($788677\times 1637$).
An exploratory procedure was written in Interactive Data 
Language (IDL), and it turned out to be faster than 
expected since convergence occurs in, typically,
less than 50 iterations. Using a 
8-core Intel Xeon 2.66\,GHz machine with 32\,GB
of RAM,
50 iterations last less than 300 minutes. 
(The access to sufficient RAM was critical, 
since the array with the spectra to be classified 
occupies some 11.6\,GB.)
Even if fast, the IDL code does not allow us 
to carry out the battery of classifications 
required to study the dependence of the 
classification on the random initialization
(\S~\ref{algorithm}). 
Fortunately, the \kmeans\ algorithm can be 
parallelized, and we developed a second 
parallel version of the code using Fortran  
and the MPI (Message Passing Interface) library. 
% One have to modify the serial 
% version of the algorithm 
% so that it distributes all the spectra
% among the different processors. 
The performance of the parallel version is good. 
The algorithm scales very well, so that adding more 
CPUs implies a near to linear reduction 
in the execution time.
A hundred  executions of the parallel code 
using a cluster of 48 Intel Xeon CPUs (2.4\,GHz) 
takes of the order of 1 hour. This figure 
outperforms the IDL code by a factor of 500. 

\begin{figure}
\includegraphics[width=0.45\textwidth]{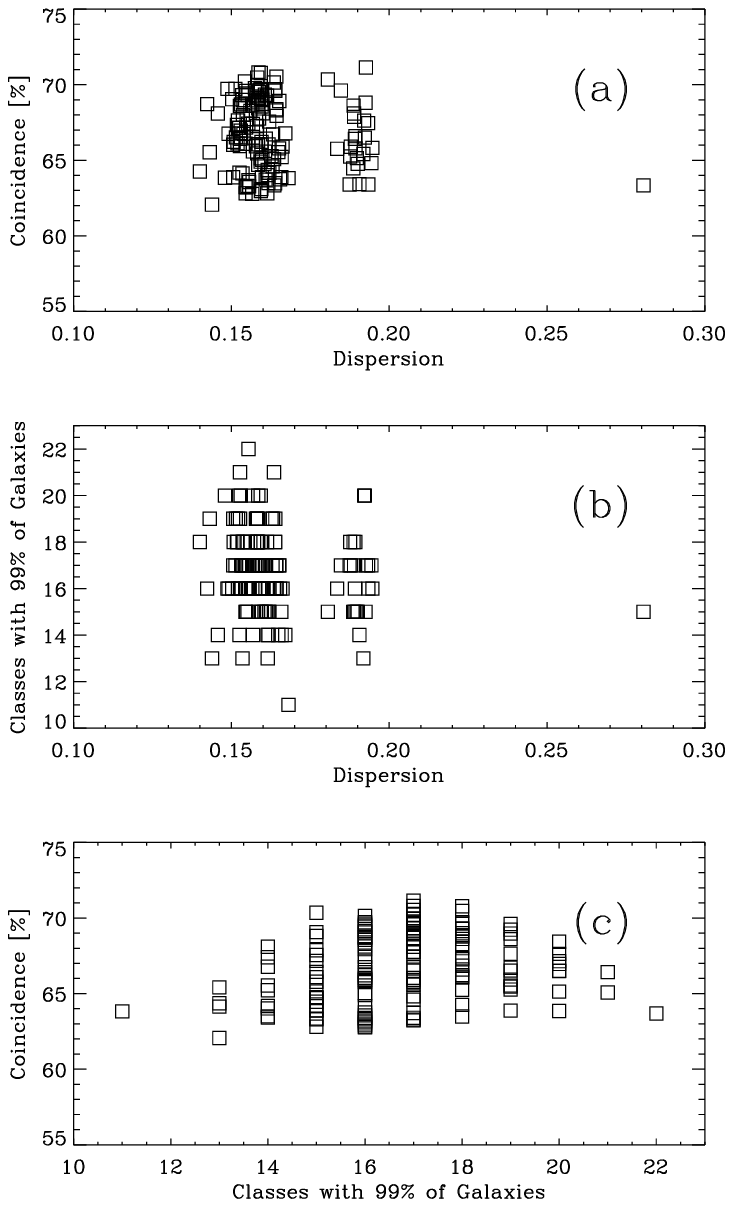}
\caption{
Scatter plots with the three parameters characterizing the 150 
different classifications from which we have drawn  
the final one in Fig.~\ref{classification}. 
(a) Percentage of galaxies common to all other 
classifications (coincidence) versus typical scatter 
of the spectra with respect to the class spectrum 
(dispersion). 
(b) Number of major classes versus dispersion.
(c) Coincidence versus number of major 
classes.
We select classifications having coincidence~$> 70\%$,
dispersion~$< 0.17$, and 17~classes or less. 
}
\label{compare_chi2}
\end{figure}

Aided with the parallel version of \kmeans ,
we carry out 150 independent classifications
of the data set. Because of the random initialization, 
each one of these classifications 
differs (\S~\ref{algorithm}). Each run of the 
algorithm groups similar spectra in clusters so,
in principle, 
all of them provide valid  classifications. 
Then the problem arises as 
to which one of these classifications is best,
i.e., which one should be chosen as {\em the} classification.
Ideally, one would like to choose a classification  
(1) with a small number of classes,
(2) being representative of all classifications,  
and (3) having small dispersion within the classes.  
Condition (1) is obvious and will not be discussed further. 
According to condition (2), we would like the classification
to be as representative as possible of any other 
classification. Condition (3) demands that the spectra
employed in deriving the classification are 
as close as possible to a class center spectrum. 
Figure~\ref{compare_chi2} shows scatter plots 
of three numerical coefficients that we devised 
to quantify the three requirements above. 
Figure~\ref{compare_chi2}a and  \ref{compare_chi2}c
include the average percentage of galaxies that 
a particular classification has in common with the other 
149 classifications -- it is just the percentage of galaxies 
in equivalent classes as defined in \S~\ref{test_qbcd}, 
and it is labeled in the figures  as {\em coincidence}. 
It  spans between 62\% and 71\%. 
The coincidence is represented in Fig.~\ref{compare_chi2}a
versus the average dispersion of the classification, which
is just the mean of 
distance between galaxies and class
centers defined in equation~(\ref{def_dispersion}). 
The dispersion admits a simple interpretation: it is the typical
difference per pixel between a spectrum in its class. 
(Because of the normalization, 
the spectra have their continuum at about one, 
therefore, dispersion 0.1 corresponds to 
differences of the order of 10\%.)
Note that there is no obvious correlation between
the two parameters, but the classifications 
seem to cluster around two dispersions, the smallest
being of the order of 0.16.
Figure~\ref{compare_chi2}b shows the scatter plot between
the number classes in a classification (classes 
altogether containing
99\% of the galaxies) and the dispersion. 
Classifications
having between 11 and 22 classes exist, 
with a typical value between 15 and 19. Again no
obvious relationship between number of classes and
dispersion is observed. 
Finally, Fig.~\ref{compare_chi2}c shows the scatter 
plot of coincidence versus number of classes. Attending to the 
three  requirements above, we select those 
classifications having
\begin{enumerate}
\item less than 18 classes,
\item coincidence larger than 70\%,
\item dispersion smaller than 0.17.
\end{enumerate}
%%%
Four classifications fulfill these requirements.
Lacking a better criterion, we choose one of 
them at random. 
The chosen classification turns out to have a 
coincidence of 70.8\%, a dispersion of 0.16, 
and it has 28 classes, but 17 of them 
contain 99\% of the galaxies. These 17 classes
are denoted in the paper as {\em major} classes. 
The spectra of the major classes
are shown in Fig.~\ref{classification}.
They have been labeled according to the $u-g$ color, 
from the reddest, \class{0}, to the bluest, 
\class{27}. By using numbers to label the classes
we are not implicitly assuming that the spectra represent
a one dimensional family. The numbers are only
tags to name the  classes.

\begin{figure}
\includegraphics[width=0.8\textwidth,angle=90]{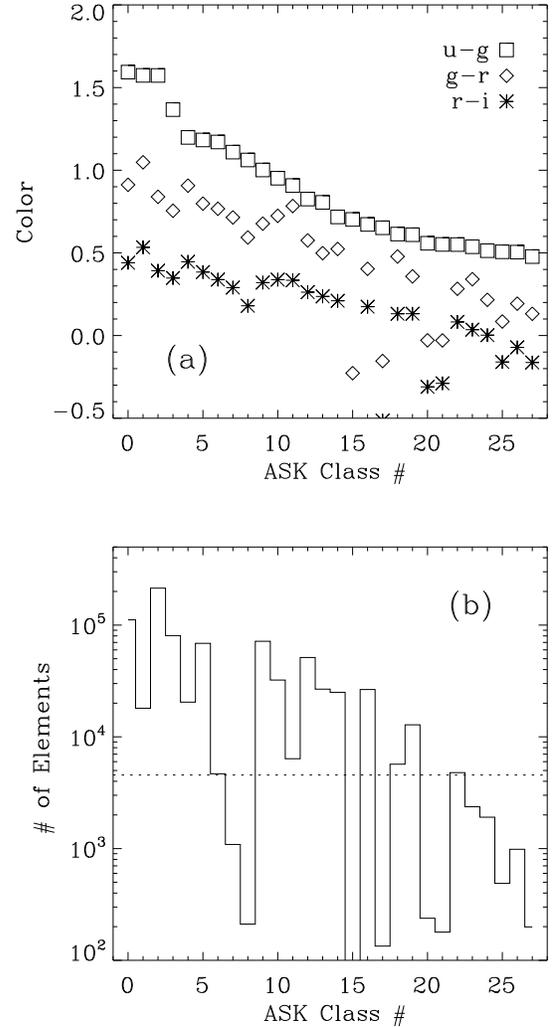}
\caption{
(a) Color versus ASK class number. Class numbers have been assigned 
according to the $u-g$ color of the template 
spectra, which explains
the monotonous decrease of this color with class number.
The larger the class number the bluer the galaxy.
Colors $g-r$ and $r-i$ are also included as indicated
by the inset.
(b) Histogram of the number of galaxies existing
in each class. The horizontal dotted line shows the
threshold that separates major classes from the rest
(i.e., classes having altogether 99\% of the classified galaxies).
The colors and number of members shown in these
figures are listed in Table~\ref{tab2}.
}
\label{sorting}
\end{figure}

\begin{figure*}
\includegraphics[width=.70\textwidth,angle=90]{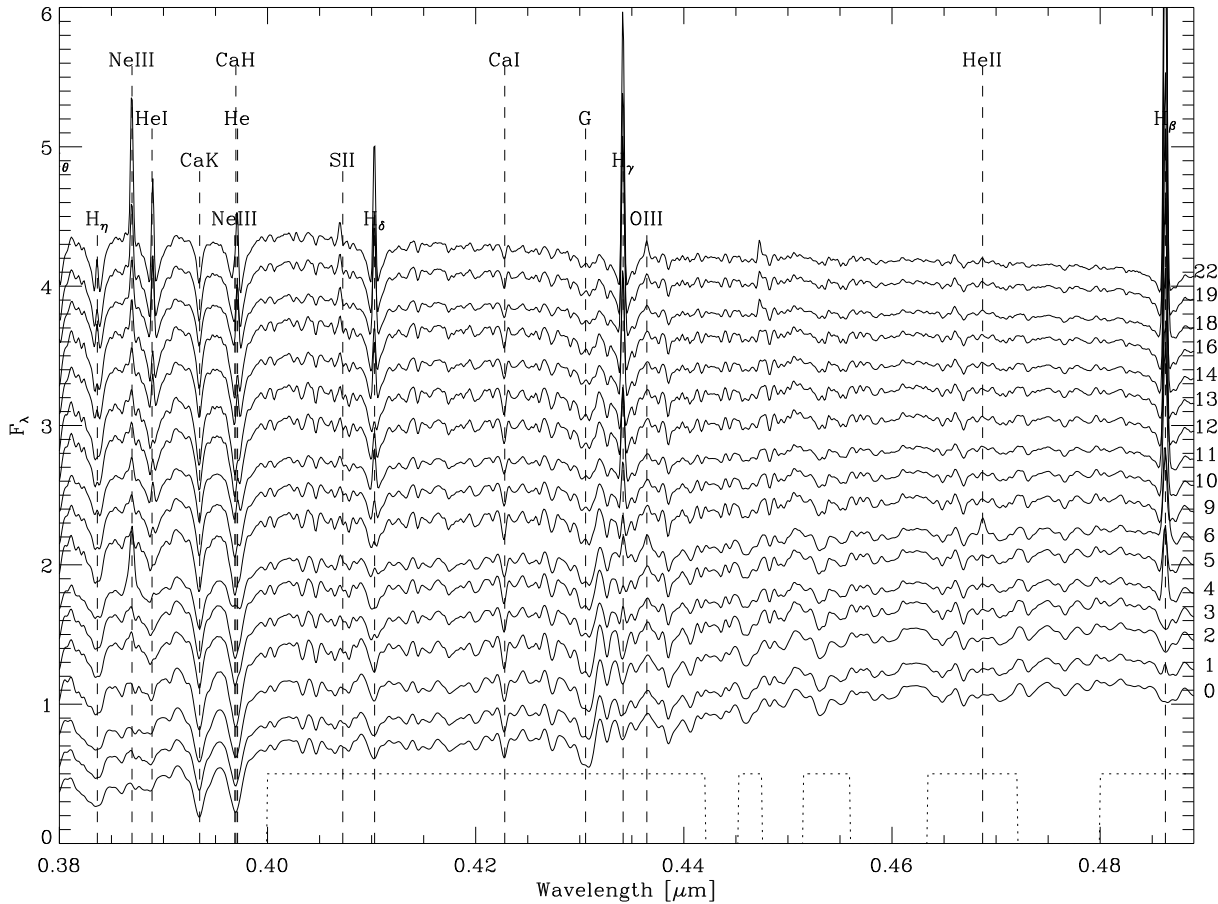}
\caption{
Zoom into the bluest wavelength range 
of Fig.~\ref{classification}.
In addition to the lines and labels in the original
figure, this plot also includes
the most prominent spectral lines 
(vertical dashed lines with labels). 
For scaling, shifting, and further details, 
see the caption of Fig.~\ref{classification}.
}
\label{classificationc}
\end{figure*}

Figure~\ref{sorting}a shows the colors
characteristic of the ASK classes.
The number of elements corresponding to each class
is included in Fig.~\ref{sorting}b.
The horizontal dotted line in this figure
indicates the threshold for major class, 
i.e., classes with a number of elements above 
this threshold contain 99\% of the classified 
galaxies. Their 
spectra are those shown in Fig.~\ref{classification}.
The main properties of all classes are summarized
in Table~\ref{tab2}.
A blow up with the bluest part of Fig.~\ref{classification}
is  included in Fig.~\ref{classificationc},
where some of the characteristic emission and 
absorption features are labeled.
Note how the spectra vary gradually with the
class number. Even the smallest 
ripples in these average spectra are real.
Upon averaging, the signal-to-noise ratio is expected
to increase as the square root of the number
of class members.
The major class with less members still has 
$\sim$\,5000 elements (Fig.~\ref{sorting}b),
which sets a lower limit to the S/N per pixel 
of $\sim 700$. 
The systematic change of global properties
along the sequence is important to constrain the
effects of noise on the number of classes
(\S~\ref{test_qbcd}). Altough noise 
artificially increases the number of classes, 
it does not change in a systematic way global 
spectral properties such as colors. 
Except perhaps at the blue 
end of the classification, the colors of classes 
vary systematically along the sequence
(Fig~\ref{sorting}; see also \S~\ref{colorscolors}), 
which discard any significant influence of the 
random pixel-to-pixel uncorrelated noise 
on the number of classes.
\begin{deluxetable*}{clccccccccr}
% \begin{deluxetable}{clccccccccr}
\tablecolumns{10}
\tablewidth{0pc}
\tablecaption{Main properties of the ASK classes}
\tablehead{
\colhead{ASK class\tablenotemark{1}}    &  
\colhead{members}   &
\colhead{$u-g$\tablenotemark{2}} &
\colhead{$g-r$\tablenotemark{2}} &
\colhead{$r-i$\tablenotemark{2}} &
\colhead{H$\beta$\tablenotemark{3}}&
\colhead{[OIII]$\lambda5007$\tablenotemark{3}}&
\colhead{H$\alpha$\tablenotemark{3}}&
\colhead{[NII]$\lambda$6583\tablenotemark{3}}&
\colhead{\% emission\tablenotemark{4}} &
\colhead{Clustering}
}
\startdata
 0$^*$& 111447& 1.59& 0.91& 0.44& -0.3&  0.7&  0.9&  2.0&  9&neutral\cr
 1$^*$&  18032& 1.57& 1.05& 0.53&0.3&  1.1&  2.9&  3.2& 29&neutral\cr
 2$^*$& 213936& 1.57& 0.84& 0.39&-0.4&  0.4&  0.1&  1.3&  2&good\cr
 3$^*$&  80530& 1.37& 0.75& 0.35&-0.7&  0.6&  0.7&  1.6&  8&neutral\cr
 4$^*$&  20456& 1.20& 0.91& 0.45&2.7&  2.3& 13.9&  8.2& 95&bad\cr
 5$^*$&  68626& 1.18& 0.80& 0.38&1.5&  1.6&  8.5&  5.3& 78&neutral\cr
 6$^*$&   4669& 1.17& 0.77& 0.34&3.5& 29.6& 14.2& 10.9& 98&good\cr
 7 $ $&   1089& 1.11& 0.71& 0.29&8.7& 84.7& 22.5& 15.2&100&good\cr
 8 $ $&    211& 1.06& 0.59& 0.18&22.8&233.1& 36.0& 15.5& 95&good\cr
 9$^*$&  71671& 1.00& 0.68& 0.32&2.4&  1.6& 12.0&  5.8& 89&neutral\cr
10$^*$&  32227& 0.95& 0.72& 0.34&5.1&  2.7& 24.2& 11.1& 99&neutral\cr
11$^*$&   6369& 0.91& 0.78& 0.33&9.7&  5.4& 42.2& 19.8&100&good\cr
12$^*$&  51314& 0.83& 0.58& 0.26&5.4&  3.0& 24.5&  9.3& 98&good\cr
13$^*$&  26705& 0.81& 0.50& 0.24&2.6&  3.0& 11.2&  3.9& 81&bad\cr
14$^*$&  25026& 0.72& 0.52& 0.21&10.2&  6.1& 45.7& 16.1& 99&good\cr
15 $ $&     68& 0.70&-0.23&-0.57&176.4&743.5&715.3& 14.1& 25&good\cr
16$^*$&  26504& 0.67& 0.40& 0.17&7.1&  8.8& 30.4&  7.1& 99&neutral\cr
17 $ $&    134& 0.65&-0.15&-0.51&161.7&630.1&549.9& 16.4& 35&good\cr
18$^*$&   5687& 0.61& 0.48& 0.13&19.5& 18.5& 83.9& 24.9&100&good\cr
19$^*$&  12808& 0.61& 0.36& 0.13&13.0& 21.6& 54.6&  9.8&100&good\cr
20 $ $&    238& 0.56&-0.03&-0.31&105.5&492.9&408.2& 17.6& 81&good\cr
21 $ $&    179& 0.55&-0.03&-0.29&97.1&461.5&356.7& 16.6& 69&good\cr
22$^*$&   4781& 0.55& 0.28& 0.08&19.8& 48.3& 82.5&  9.7&100&good\cr
23 $ $&   2366& 0.54& 0.34& 0.04&31.5& 61.1&130.7& 22.8& 89&neutral\cr
24 $ $&   1910& 0.51& 0.22& 0.00&34.5&106.6&148.5& 12.9& 98&good\cr
25 $ $&    488& 0.51& 0.08&-0.16&72.3&253.8&302.7& 18.5& 94&good\cr
26 $ $&    986& 0.50& 0.19&-0.07&51.7&159.9&219.3& 19.6&100&good\cr
27 $ $&    199& 0.48& 0.13&-0.16&67.4&230.7&278.8& 19.7& 85&good
\enddata
\tablenotetext{1}{The asterisks denote
major classes, i.e., those that 
altogether include 99\% of the galaxies.}  
\tablenotetext{2}{The colors have been
computed from the template spectra 
using the appropriate SDSS bandpasses.}
\tablenotetext{3}{Equivalent width in the
template spectra given in \AA .
Negative implies line in absorption.}
\tablenotetext{4}{Percentage of 
galaxies in the class with 
H$\beta$ in emission according to the SDSS/DR7 catalog.}
% \tablecomments{}
\label{tab2}
\end{deluxetable*}
% \end{deluxetable}

As we explained in the first paragraph of the section,
the full set of galaxies was split into two parts.
The low redshift part has been used to derive 
spectral classes, which automatically leads to its
classification as explained above.
Based on these classes, and using the procedure 
developed in \S~\ref{assigning}, 
we extended the classification to the high redshift 
subset. The use of the same classes is an
assumption which, however, seems to be secure since
the  
properties of the galaxies thus classified do not
show any systematic difference with respect to the
low redshift subset (see \S~\ref{ask_vs_redshift}).  
Moreover, the procedure in \S~\ref{assigning} has been 
also applied to the low redshift part, already
classified by \kmeans . It provides qualities
for all classes, which permits identifying 
borderline galaxies and outliers, and it 
allows us deriving
physical parameters by interpolation (\S~\ref{applications}).

\begin{figure*}
\includegraphics[width=0.7\textwidth,angle=90]{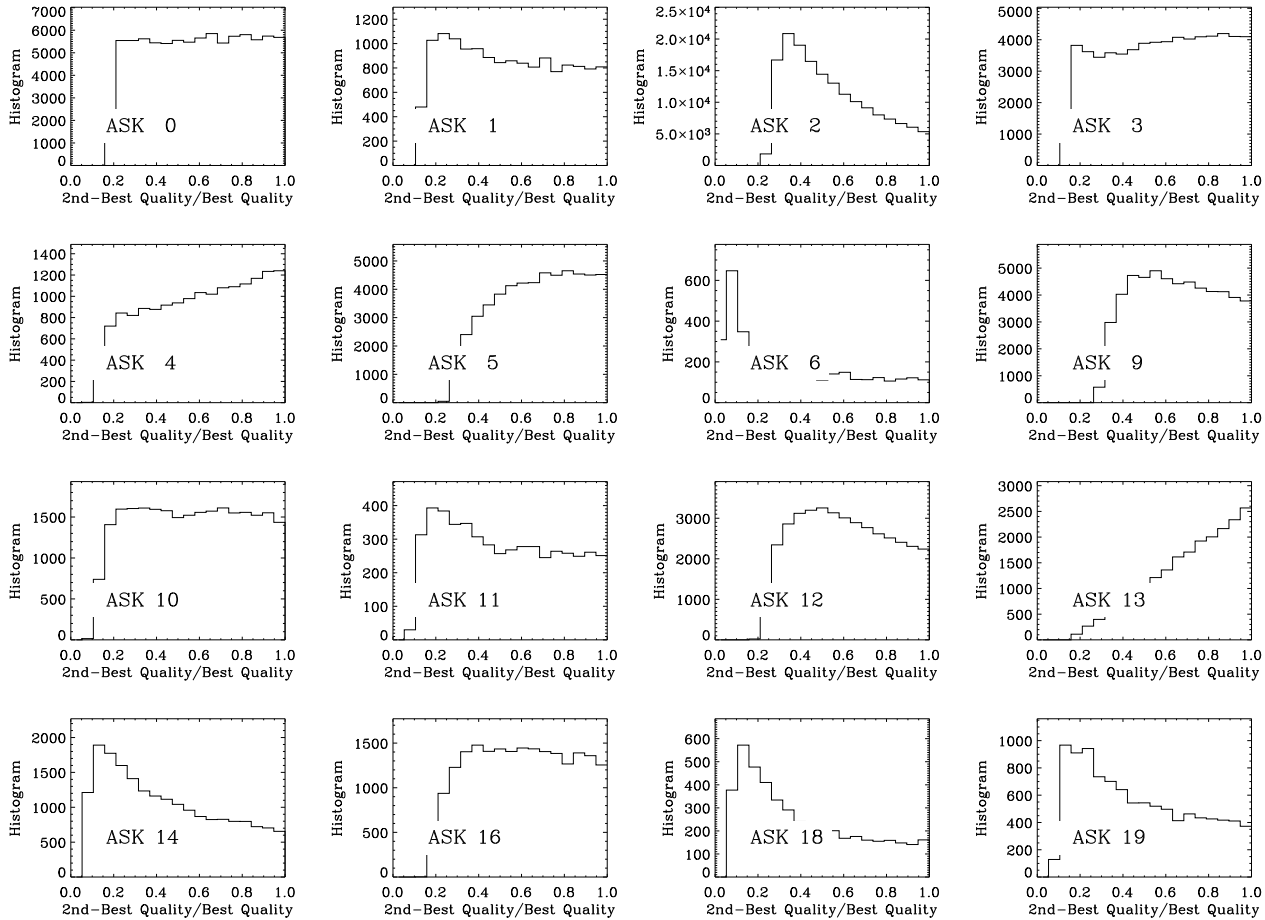}
\caption{Histograms of the ratio between the 
best and the 2nd-best qualities for the major
classes. Those classes corresponding to proper 
clusters in the 1637-dimensional classification 
space should have a distribution of ratios peaking
away from one (e.g., ASK~2).
Only galaxies whose best quality is 
larger than 0.2 have been considered.
}
\label{membership}
\end{figure*}

The success of \kmeans\ 
does not imply the existence of well defined 
clusters in the  1637-dimensional classification space. 
As we discuss above, the separation between classes is not
sharp. Galaxies are often close to the borders, 
which explains the variability between different 
realizations of the classification  
(\S~\ref{test_qbcd}). 
The presence of many borderline galaxies
seems to imply a rather continuous distribution 
of points in the classification space, 
that \kmeans\ astutely partakes assuring the 
elements of each class to be similar.
Generally speaking, classes should not
be associated with true clusters in the classification
space. However, some of the classes  seem to represent genuine 
clusters as judged from the distribution of 
{\em qualities}. Qualities were 
introduced in \S~\ref{assigning} to 
characterize the membership of each galaxy to 
the classes. Galaxies next to class borders have 
similar best quality and 2nd-best quality.
Therefore, if the galaxies in a class are of this
kind, then the class cannot portray a clear cluster. 
Conversely, classes corresponding to well defined 
clusters have their members separated from the other
classes, i.e., their galaxies tend to have a 
best quality larger than the 2nd-best 
quality. This condition is met by 
some of the classes, indicating clustering.
Thus we use the ratio between best and 
2nd-best qualities to assign a degree
of grouping to the different classes.   
Figure~\ref{membership} shows histograms
of the ratio between the best quality and 
the 2nd best quality for the major classes. 
Some of the histograms have a clear peak at a
ratio significantly smaller than one, implying
clustering  (e.g., ASK~2). Other histograms
present a flat distribution (e.g., ASK~0),
whereas a minority of classes show 
most of their members having similar best and
2nd-best qualities (e.g., ASK~13).
Attending to the shape of these histograms,
the clustering of each class has been 
labeled as good, neutral or 
bad; see the last column of Table~\ref{tab2}. 
Note that the clustering tends to be good or neutral 
rather than bad.

%
%%%%%%%%%%%%%%%%%%%
%
\section{Relationship between ASK class 
and PCA classification}\label{pca_class}

\begin{figure*}
\includegraphics[width=0.7\textwidth,angle=90]{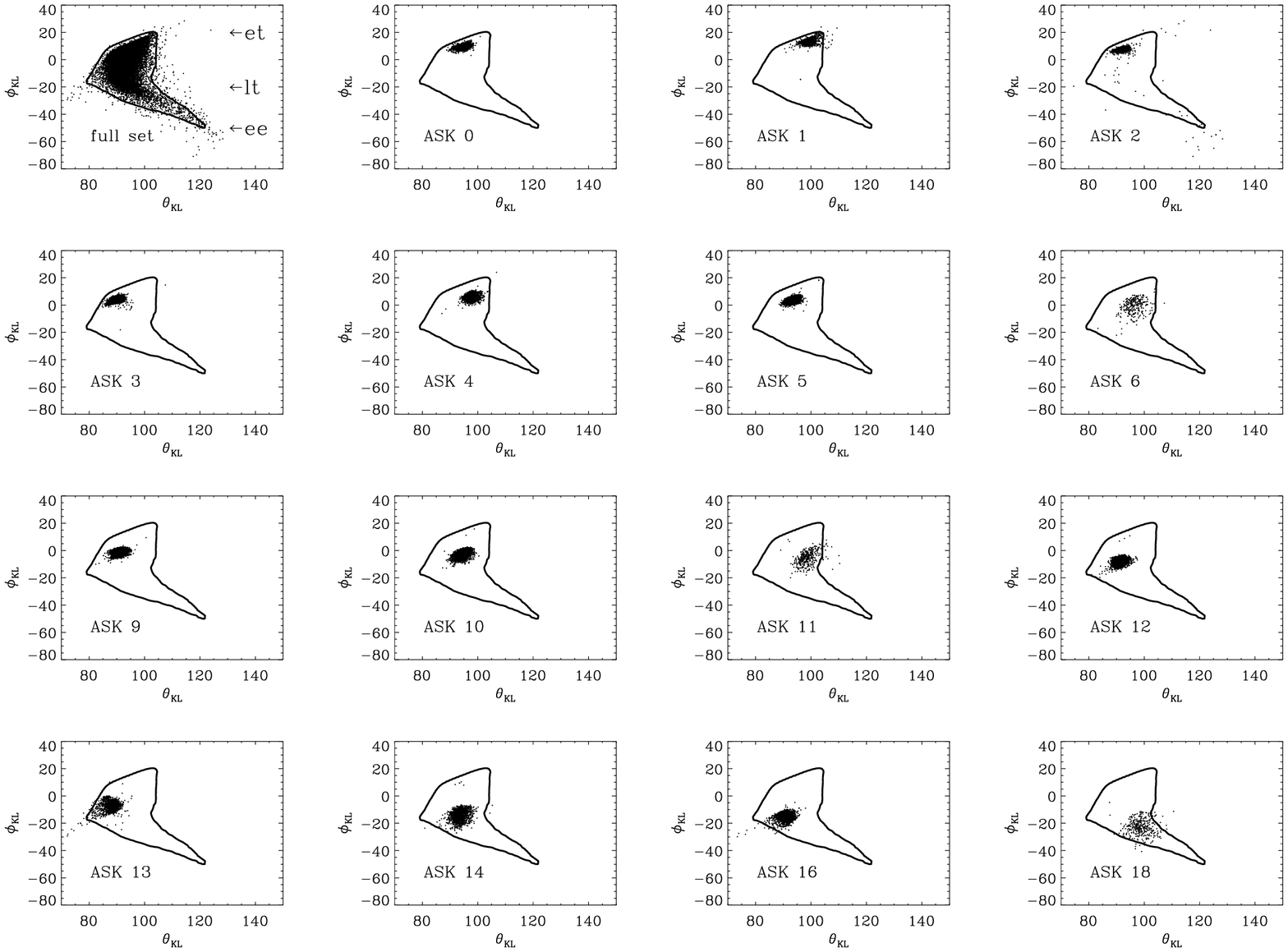}
\caption{PCA diagnostic plot for   
50000 randomly chosen galaxies in the SDSS/DR7. The full set is 
included in the top left panel,  where we also 
mark the location 
of the early~type (et), late~type (lt)
and extreme~emission (ee) galaxies according 
to the separation in \citet{yip04}. 
The other plots show the different
major ASK classes individually (see the insets).
Note how the ASK classes occupy well defined 
places in the PCA diagnostic plot.
Given the ASK class of a galaxy, one can predict 
its location in the PCA plane. The opposite does 
not hold in general.  All plots include 
the same contour indicating the boundaries of the  
full distribution. 
The two mixing angles $\theta_{\rm KL}$
and $\phi_{\rm KL}$ are given in degrees.
Major classes 19 and 22 are not shown because
they do not fit into the figure, but they
follow the sequence.
}
\label{ask_vs_pca}
\end{figure*}

SDSS/DR7 already provides a spectral classification based
on PCA, which is a linear expansion of each spectrum in terms of a 
small number of eigenspectra (\S~\ref{intro}).  
The eigenspectra for the 
SDSS expansion were derived from a subset
of approximately 200000 galaxies, as explained by
\citet{yip04}. Following \citet{con95} and others, 
\citet{yip04} use a diagnostic plot to separate 
spectral classes based on the three first eigenvalues, 
$a_1, a_2$ and $a_3$. Extreme emission galaxies, 
early type galaxies, and late 
type galaxies can be distinguished in the 
$\phi_{\rm KL}$ versus $\theta_{\rm KL}$ plane,
where the two mixing angles are defined as 
\begin{eqnarray}
\phi_{\rm KL}=&\arctan(a_2/a_1),\cr
\cr
\theta_{\rm KL}=&\arccos(a_3).
\end{eqnarray}
Figure ~\ref{ask_vs_pca} shows this diagnostic plot
for all SDSS/DR7 galaxies, and for the major ASK classes 
separately. 
We have  used the PCA eigenvalues directly provided 
by SDSS/DR7. The ASK classes occupy well defined
places in the PCA diagnostic plot, which implies
that the PCA classification and the ASK classification
are consistent. Given the ASK class of a galaxy, 
one can predict its location in the PCA plane. 
The opposite does not hold since 
some ASK classes overlap in the 
PCA diagnostic plot (cf. ASK~9 and ASK~10).
The ASK classification is 
more refined; it simply
includes more classes than PCA,
therefore, the two classifications are consistent
but not equivalent. 
The location of early type galaxies, 
late type galaxies, and extreme emission galaxies
made by \citet[][]{yip04} is also
included in Fig.~\ref{ask_vs_pca} (top left 
panel, symbols et, lt, and ee, respectively). One can 
see how this rough PCA based separation is also 
consistent with the ASK classes. There is  
a systematic trend to go from the location
of the early types to the late types as the 
ASK class number increases. This behavior 
coincides with the trend to be derived from the
morphological classification in 
\S~\ref{ask_class_vs_morph}.
The region of extreme emission galaxies
deserves a separate comment. Note that the galaxies
appearing in this region do not show
up among the galaxies in the major classes
included in Fig.~\ref{ask_vs_pca}. These
extreme galaxies belong to the minor classes
with high ASK class number (not shown), i.e.,
the bluest among the ASK classes.
The points clearly outside the contour 
in  Fig.~\ref{ask_vs_pca} are partly included in 
the ASK classes next to them, 
and partly in additional minor classes (not shown). 
ASK~2 seems to be the only exception. It includes
a few galaxies in the extreme emission region,
and we have not been able to pin down the cause.
However, the fact that ASK~2 shows more outliers
then other classes is probably an artifact due to 
ASK~2 being the most common class (Table~\ref{tab2}).
If all classes include a similar fraction of outliers, 
they will be more conspicuous in scatter plots of ASK~2. 

In short, although ASK is 
more refined,
ASK and PCA seem to agree with small
internal scattering. 
Moreover, the scatter between these two 
purely spectroscopic classifications is 
much smaller than the scatter in the 
ASK versus morphological classification analyzed
in the next section.

%
%%%%%%%%%%%%%%%%%%%%%%%%%%%%%
%
\section{Relationship between 
ASK class and Hubble type}\label{ask_class_vs_morph}

The morphological type of a galaxy (Hubble type) 
is closely related to its spectrum, 
a relationship known for long (see \S~\ref{intro}).
The analysis of such relationship in the case
of ASK is mandatory, 
and we will do it in a follow-up work where 
the morphology of a large number of galaxies
is derived automatically 
(\S~\ref{conclusions}). However, in order to show 
the consistency of the ASK classification, 
we include here a preamble based on
a limited number of galaxies which shows how 
early types are associated  with small ASK numbers, 
and vice-versa.

\begin{figure}
\includegraphics[width=0.45\textwidth,angle=0]{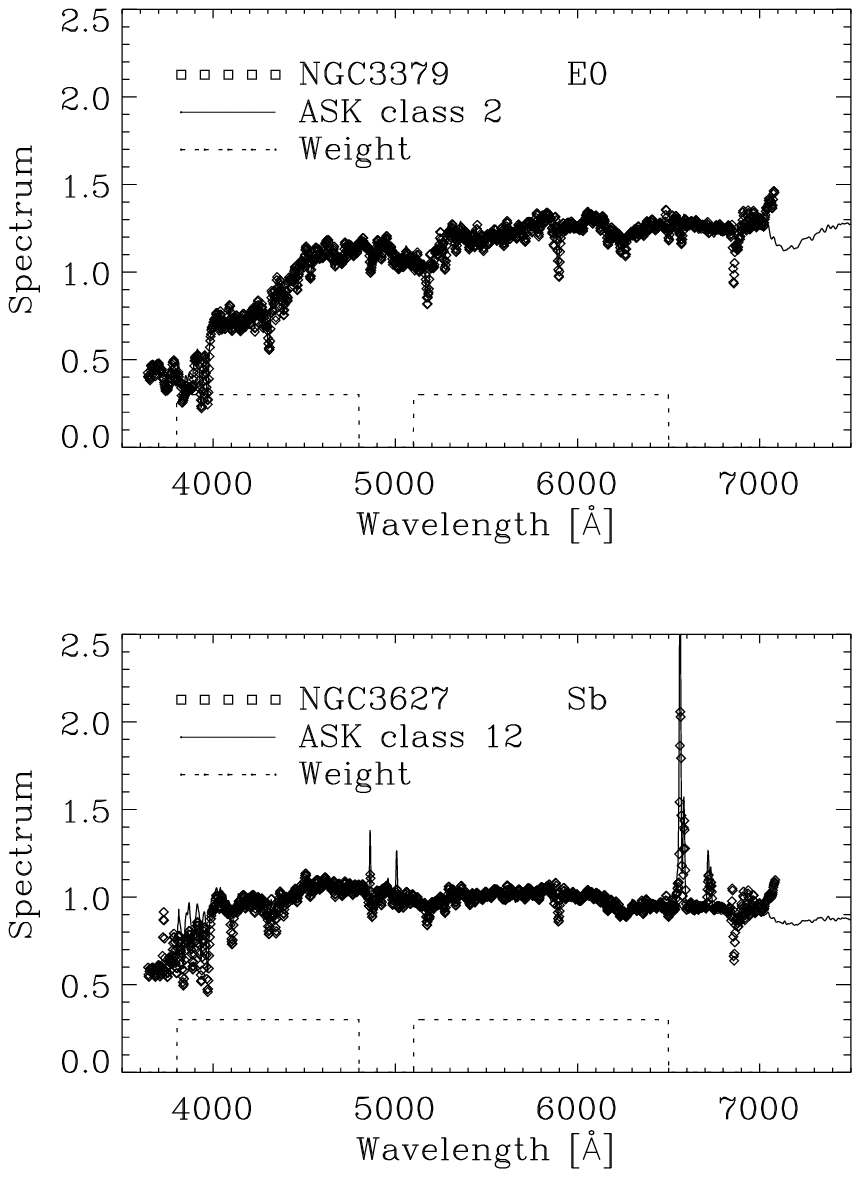}
\caption{
Two representative examples of the fits 
between ASK class spectra and 
galaxies in the atlas by \citet{ken92}.
Galaxy names, Hubble types, and ASK numbers are
included in the insets. The dotted lines correspond
to the weights used for fitting -- wavelengths
where the weight is zero have been ignored. 
Wavelengths are given in \AA .
Symbols and solid lines correspond to \citeauthor{ken92}
spectra and ASK class spectra, respectively.
}
\label{assign_kennicutt_other}
\end{figure}

\begin{figure*}
\centering
\includegraphics[width=0.8\textwidth]{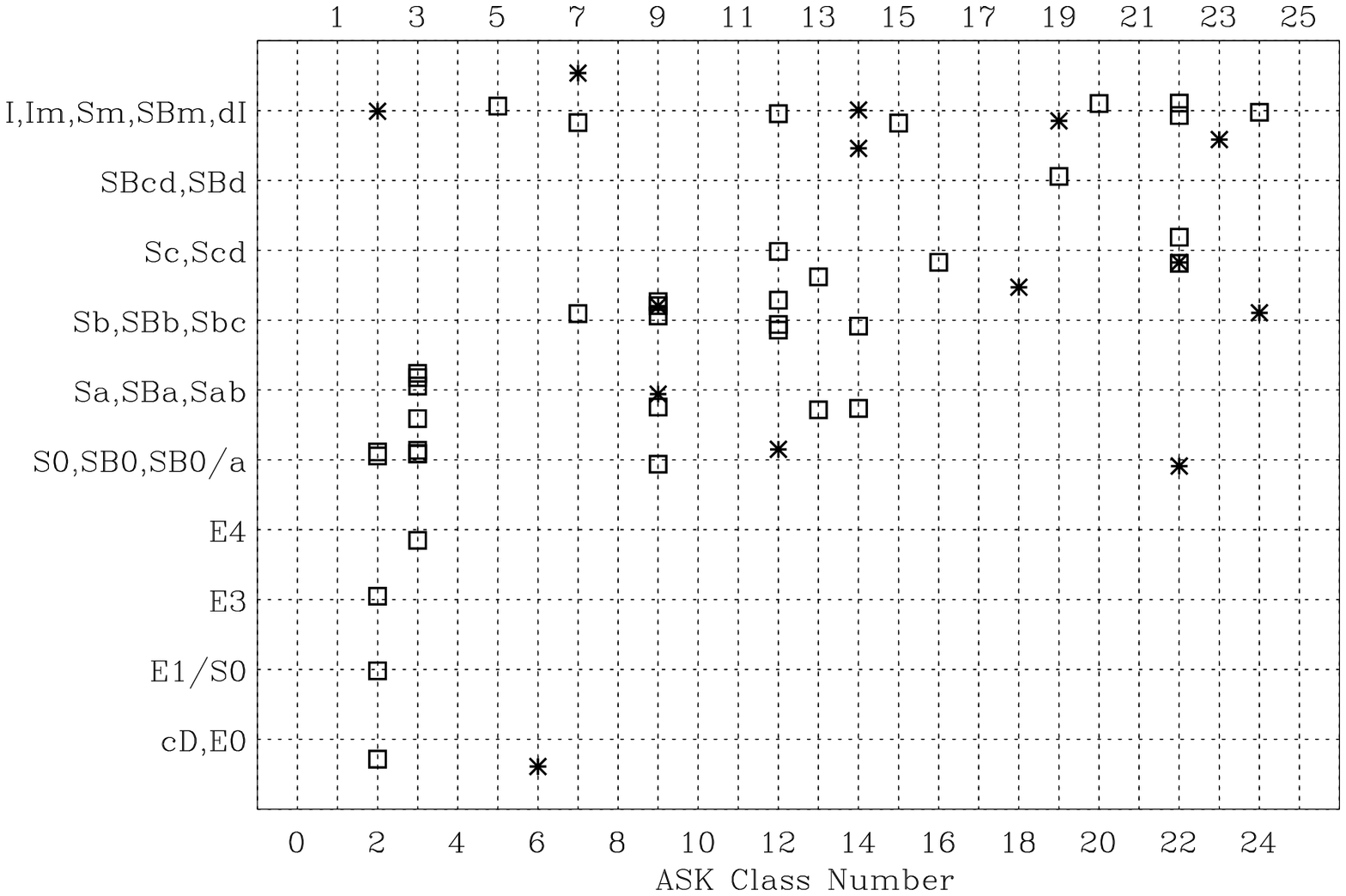}
\caption{
Scatter plot of Hubble type versus ASK class for the
galaxies in the atlas by \citet{ken92}. 
The plot contains 53 out of the 55 galaxies in the atlas --
Mrk\,3 has no spectrum in the electronic catalog, and NGC\,3303
belongs to an undefined class since it is undergoing a major merger.
Note the clear trend for the small ASK 
class numbers to be associated with early types,
and vice-versa. The trend is even more clear if one
ignores those galaxies classified as peculiar
by \citet{ken92} (the asterisks). However,
one cannot ignore the scatter -- there is no
one-to-one relationship between
spectroscopic class and morphological class. 
In order to avoid the overlapping of the galaxies
with the same Hubble type and ASK class, we
have added a small artificial random vertical 
shift to all points.
}
\label{assign_kennicutt}
\end{figure*}

We have ASK-classified the galaxies 
in the spectral atlas of \citet{ken92}. He provides 
spatially integrated spectra (from 3650\,\AA\
to 7100\,\AA ) of a set of 55 nearby galaxies 
with known Hubble types. The set contains all 
Hubble types, from gigant ellipticals (cD, NGC\,1275) 
to dwarf irregulars (dI, Mkr\,35). We assign each 
galaxy to the ASK class whose spectrum is closest 
to the galaxy spectrum as
explained in  \S~\ref{assigning}.
The match between ASK template
spectra and \citeauthor{ken92} spectra
is illustrated in Fig.~\ref{assign_kennicutt_other},
which contains representative
spectra of an early type galaxy and 
a late type galaxy.  These particular 
fits ignore the spectral regions with emission lines 
(see the weights shown as a dotted line in 
the figures). 
The scatter plot of the assignation is shown
in  Fig.~\ref{assign_kennicutt}. It displays
the Hubble type given by \citet{ken92}
versus the ASK class for the
galaxies in the atlas. (Actually, for 53 out of 
the 55 original galaxies,
since Mrk\,3  is not in the electronic 
catalog, and NGC\,3303 has no clear Hubble type --
it presents two nuclei undergoing a major merger.)
Note the clear trend for the small ASK 
numbers to be associated with early types
and vice-versa. The dividing line between 
early types (E,\dots S0) and late types  
(Sa, SBa,\dots I) seems to be about 
\class{6}, so that numbers smaller than
this limit correspond to early types.
The trend is even more clear if one
ignores those galaxies classified as peculiar
by \citet{ken92}, which are shown in the figure as asterisks. 
However, one cannot ignore the large scatter in the
figure -- there is no one-to-one relationship between
spectroscopic class and morphological class. 
The conclusion that there is a general trend
with large scatter is very much in the vein
of all previous studies comparing spectroscopic 
and morphological classifications 
\citep[e.g.,][and \S~\ref{intro.intro}]{zar95,con06}.
Actually, the relationship between morphological
and spectroscopic types gets fuzzier with 
increasing lookback time, and perhaps it 
disappears in the early Universe  \citep{con06,hue09}.

We have repeated the above exercise using the 
eye-bold morphological classification presented
by \citet{fuk07}. It is based on bright galaxies 
in a north equatorial stripe from SDSS/DR3,
visually classified by  three different observers
based on $g$-band images.
The catalog contains 2253 galaxies, 
but only 1866 targets have spectra 
and so overlap with our classification.
We have also used this set to compare 
the morphological Hubble type
and the ASK classification, with results
similar to those for \citeauthor{ken92} galaxies.
There is a global trend with significant scatter.
The size of the set allowed us to
discard several observational bias that may 
cause the scatter. It is not due to misclassifications.
The scatter is not reduced upon using only high 
quality ASK class determinations 
(quality $> 0.8$; \S~\ref{assigning}),
or when Im and peculiar galaxies are excluded
from the sample. The scatter remains considering only 
small galaxies contained within the spectroscopic 
fiber ($<$ 1.5\arcsec\ effective 
radius). The last test assures that the scatter 
is not produced by large spirals
misclassified because the SDSS spectrum
just samples their (red) bulge.
 
%
%%%%%%%%%%%%%%
%
\section{ASK classes and the bimodal color distribution}
\label{colorscolors}

The colors of the galaxies follow a bimodal distribution
\citep[e.g., ][]{str01,bal04,bal04b}, with a red population 
(the red sequence), a blue population (the blue cloud),
and  the so-called {\em green valley} in between
\citep[e.g.,][]{sal07}. The two main populations are believed 
to represent passively evolving red galaxies and blue
star-forming galaxies, with  galaxies in transition
forming the green valley. As we explain in \S~\ref{intro},
this work was partly triggered by the ability 
of \kmeans\  to distinguish  
green valley  spectra \citep{san09}. 
Therefore, we found it necessary to discuss the 
location of the ASK~classes in a plot
where the red and the blue populations
show up separately \cite[e.g.,][]{ber00,str01}.

\begin{figure*}
\includegraphics[width=0.7\textwidth,angle=90.]{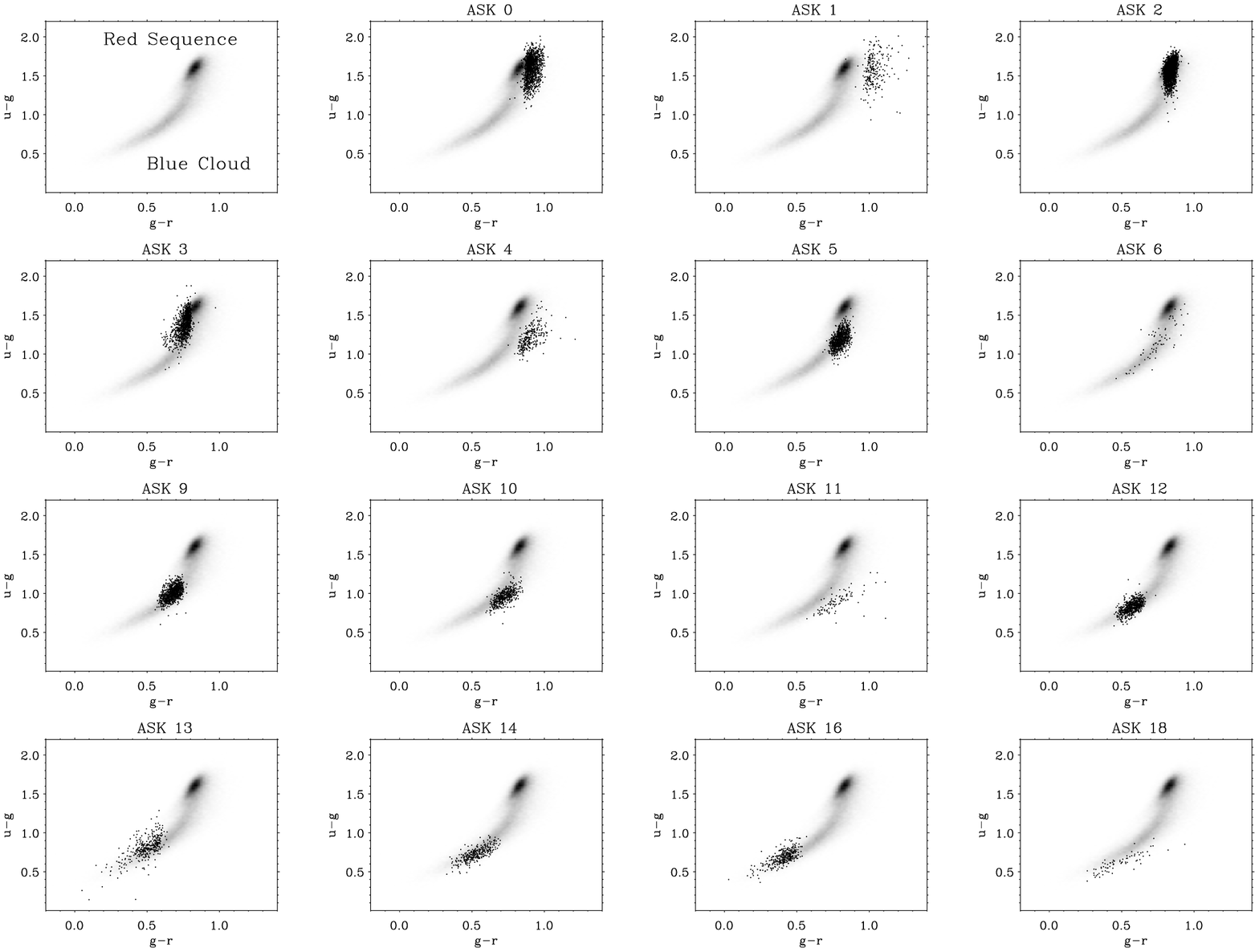}
\caption{
Plots of
$u-g$ versus  $g-r$ for the galaxies belonging
the major ASK classes. The top left panel contains 
an image with the 2-dimensional
distribution of colors in the
full SDSS/DR7. 
The remaining panels show the individual 
classes separately, as indicated in the labels,
together with the 2-dimensional histogram for reference.
Major classes 19 and 22 are not shown because
they do not fit into the figure, but they
follow the trend.
}
\label{classificationh}
\end{figure*}

Figure~\ref{classificationh} (top left panel) 
shows the distribution of all SDSS/DR7 galaxies 
in a $u-g$ versus  $g-r$ plot. The image represents
the 2-dimensional histogram  of 
the distribution of colors. 
The concentrated spot at $g-r\simeq 0.8$ and 
$u-g\simeq 1.7$ corresponds to the red sequence. 
The blue cloud appears in this representation 
as an extended tail. The other panels in 
Fig.~\ref{classificationh} show the different classes 
separately, and they all include the 2-dimensional histogram 
for reference.
An inspection to Fig.~\ref{classificationh}
reveals a number of properties. First, 
the ASK~classification separates galaxies into 
well located positions of the color-color plane. 
The red cloud is characterized 
by the most numerous class, \class{2}. 
\class{0} and \class{1} also belong to the red sequence,
but to its outskirts. The blue cloud is split
into several classes, starting with 
\class{9} and continuing with higher  
ASK classes. In between these two groups, 
\class{3}, \class{5} and \class{6} populate the green valley
-- \class{4} seems to be made of outliers of the 
main relationhip. As we originally presumed, 
the ASK classification separates galaxies in 
colors with a finesse to 
automatically pinpoint classes in the green valley.
An in-depth analysis of the galaxies in these classes
will be carried out in a follow-up work 
(see \S~\ref{conclusions}). 

%%%%%%%%%%%%%%%
\section{Relationship between ASK class and AGN activity}\label{ask_vs_agn}

We have studied the position of our classes on 
the BPT diagram  \citep[named after][]{bal81}, which is
commonly 
used to separate Active Galactic Nucleus (AGN) activity 
from normal star formation activity in galaxies with emission lines 
\citep[e.g.,][]{kau03}.
The diagnostic diagram consists of a scatter
plot of the ratio of fluxes 
[OIII]$\lambda5007/{\rm H}\beta$
versus [NII]$\lambda6583/{\rm H}\alpha$. The two pairs
of emission lines are so close in wavelength that 
the BPT diagram is almost insensitive to extinction 
and other systematic photometric miscalibrations.
\begin{figure*}
\includegraphics[width=0.7\textwidth,angle=90]{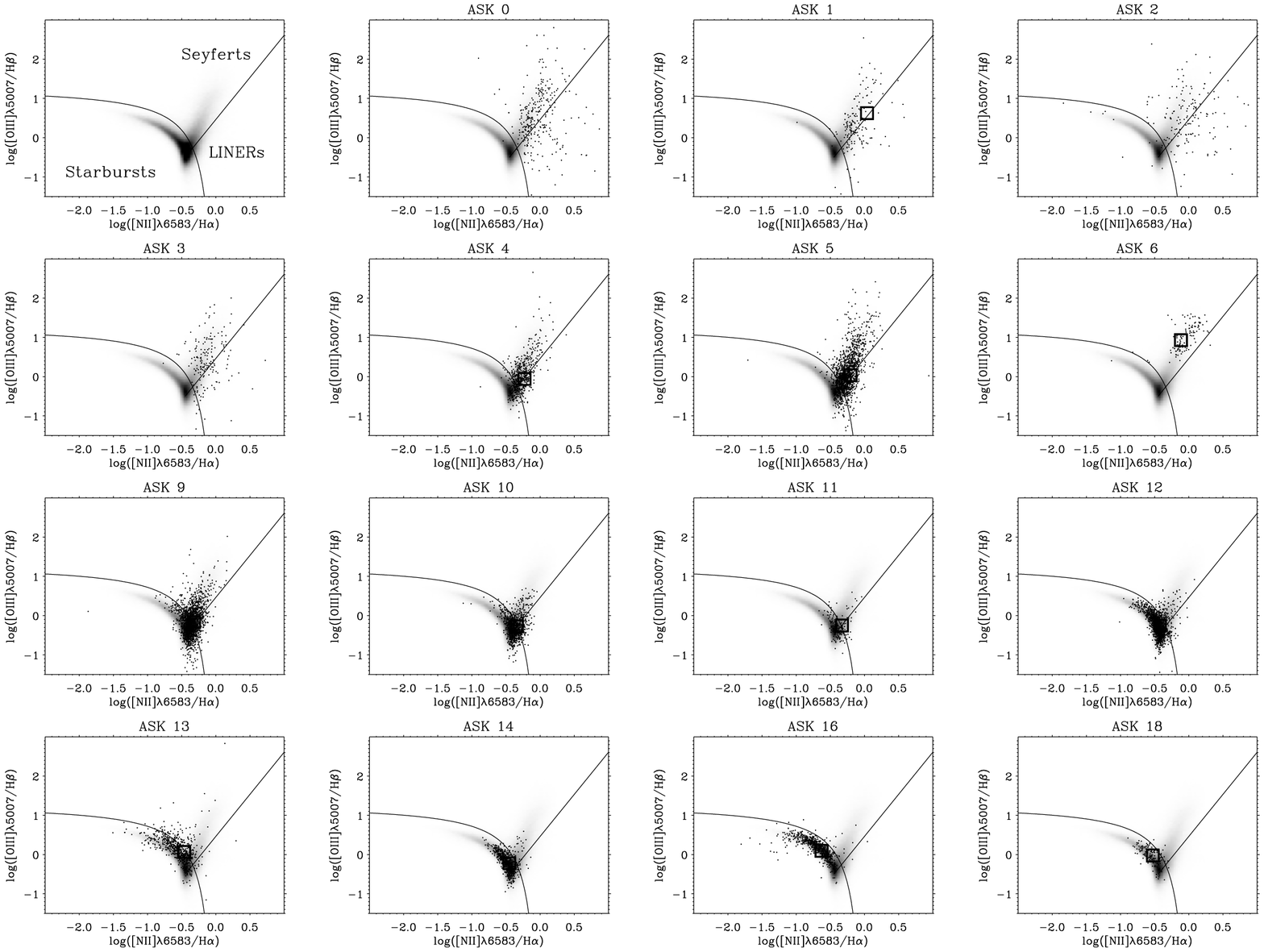}
\caption{
BPT diagrams for the full set of galaxies with emission
lines (top left panel), as well as for the galaxies 
belonging to the major ASK classes (remaining panels).
The full set is shown as an image of the
2-dimensional histogram. This image is repeated
in the rest of the panels for reference.
We represent scatter plots for 10000 individual
galaxies randomly drawn from the full SDSS/DR7
pool.
The curved solid line separates star-forming galaxies (below
the line) and AGNs (above the line).
In addition, the straight solid line in the region
of AGNs separates Seyfert galaxies and LINERs.
The plots also include a box symbol for the 
class template spectrum when it contains the 
required emission lines. 
These boxes are often buried within 
the cloud of dots. 
Major classes 19 and 22 are not shown because
they do not fit into the figure, but they
follow the sequence.
}
\label{BPT1}
\end{figure*}
Figure~\ref{BPT1}, top left panel,
contains the BPT diagram for the full set of galaxies 
with emission lines. The fluxes of the lines
have been directly taken from SDSS/DR7.
The figure includes a curved solid
line dividing star-forming galaxies (below the line)
and AGNs (above the line). This separation was worked
out by \citet[][]{kau03}, where they also
distinguish between different types of AGNs. The
straight line separates the regions occupied by Seyfert galaxies, 
and LINERs, as indicated by the insets.
The figure also shows BPT plots for the galaxies 
belonging to the individual ASK classes -- see the
class in the label on top of each plot. 
The panels for the classes include box symbols at 
the positions where class template spectra show up. 
They are barely visible because they 
always appear in the center of the cloud of points
corresponding to the individual galaxies. (ASK 0, 2, and 3
do not have such boxes since 
they present H$\beta$ in absoption; see Table~\ref{tab2}.)

Only a small fraction of red galaxies have emission lines
that can be used to place them in the BPT diagram 
(2\% for \class{2}; see Table~\ref{tab2}), however,
when they do, their emission corresponds 
to AGN activity (Fig.~\ref{BPT1}, \class{0--2}).
This result is very much in agreement with 
the current views that host galaxies of AGNs are 
preferentially early type galaxies 
\citep[][and references therein]{kau03}.
Conversely, blue galaxies correspond to star-forming
galaxies, with little sign (if any) of undergoing 
AGN activity (from \class{9} on; see 
Fig.~\ref{BPT1}).
The galaxies that seems to be 
in the green valley (\class{3}, \class{5} and \class{6}; 
\S~\ref{colorscolors}),
also appear in the BPT diagrams in the region of AGNs.
This is again consistent with the current wisdom
that AGN activity quenches star formation, and so it
may be responsible for the transit of galaxies
across the green valley \citep[][]{sch07,sch09}
The case of \class{6} deserves special attention.
According to the position in the BPT diagram,   
there is little doubt that it is formed by Seyfert 
galaxies. This is consistent with the shape of H$\alpha$
in the class template spectrum,
with very broad wings that extend up to 
2000~km\,s$^{-1}$. Moreover the only galaxy in
the catalog by \citet{ken92} classified
as \class{6} is a well known cD elliptical with a 
Seyfert nucleus (NGC~1275; see Fig.~\ref{assign_kennicutt}).

%%%%%%
%
%
\section{Relationship between ASK class and redshift}\label{ask_vs_redshift}

\begin{figure}
\includegraphics[width=0.45\textwidth,angle=90]{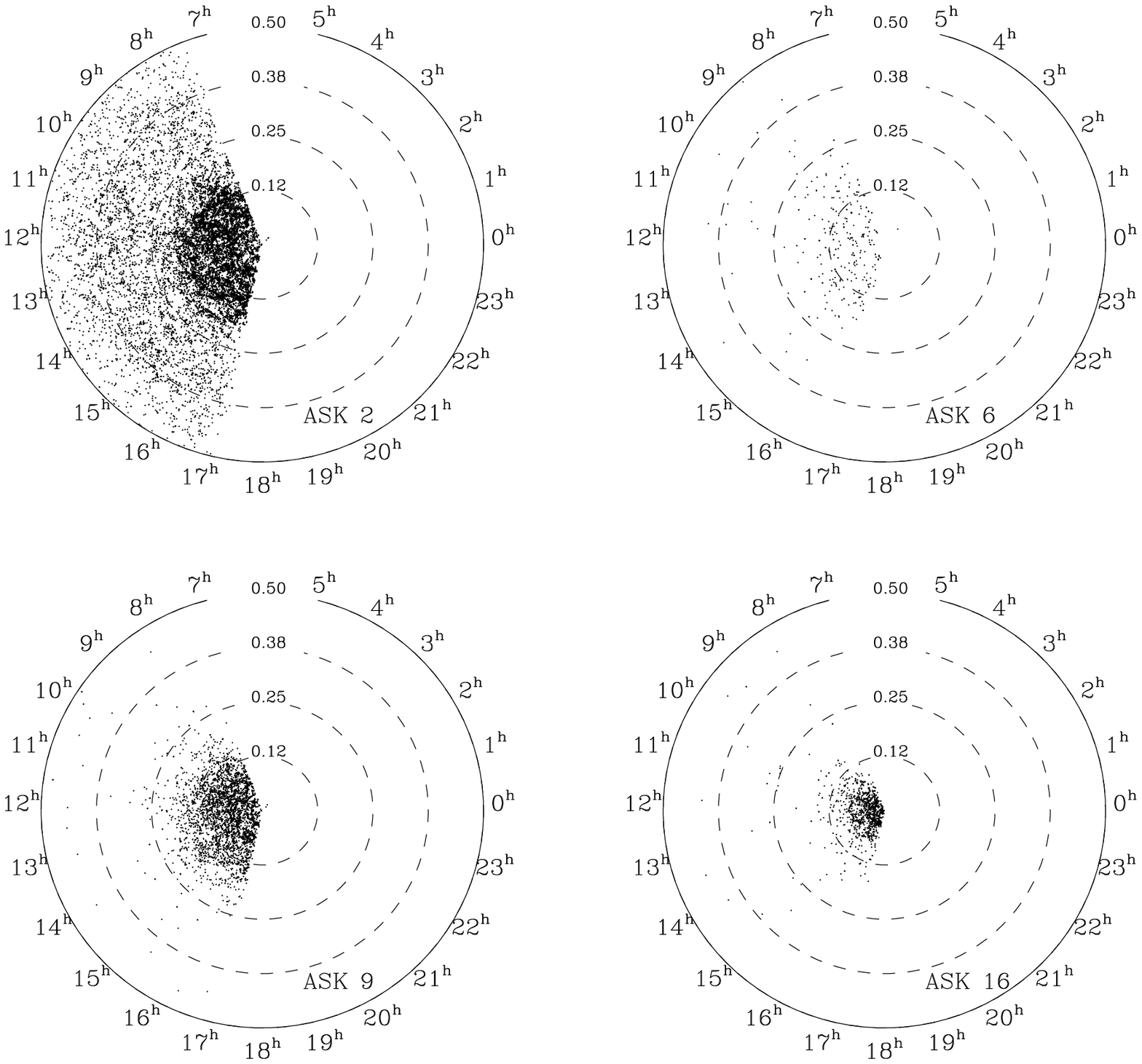}
\caption{
Cone diagrams  of four representative
classes: a red galaxy class (\class{2}), 
an AGN class (\class{6}), 
and two classes of blue 
galaxies (\class{9} and \class{16}).
Cone diagrams  are polar plots 
where radius is redshift and 
azimuth is right ascension. 
In this case $35^\circ\le$ declination $\le 45^\circ$, and 
redshift $\le 0.5$ (see the labels on the rings).
We are representing only a small fraction of 
all galaxies to avoid cluttering. 
}
\label{cone_diagram1}
\end{figure}

Cone diagrams  (or {\em pie plots}) are polar 
plots where radius is redshift and 
azimuth is right ascension \citep[e.g.,][]{fol99}.
Figure~\ref{cone_diagram1} shows 
cone diagrams  for four representative classes, i.e., 
a red galaxy class (\class{2}), an AGN class 
(\class{6}),  and two blue galaxy classes 
(\class{9} and \class{16}). 
The range of declinations is limited
between $35^\circ$ and $45^\circ$. 
From Fig.~\ref{cone_diagram1} and similar plots 
considering other classes, other range of redshifts, 
and other declinations, we draw the following conclusions.
ASK 0, 1, 2, and 3 are observed at higher redshifts 
than the rest of the classes. This effect is 
partly due to the luminous red galaxy (LRG) extension 
of the main SDSS spectroscopic sample  
\citep[][]{eis01}.
The LRG search has been designed
to detect passively evolving red galaxies,
and it includes galaxies fainter than the main 
flux-limited portion of the SDSS galaxy 
spectroscopic sample 
(down to  $r\simeq$ 19.5, rather than 
the regular cutoff at $r\simeq$ 17.8). 
However, a part of the separation 
in redshift between blue galaxies and red 
galaxies is believed to be real. Dwarf 
galaxies cannot be observed at high
redshift, but
dwarf field galaxies tend to be 
starforming \citep[e.g.,][]{hea04}, 
and so included among the bluest ASK 
classes. It is therefore understandable 
why  blue ASK classes are biases 
towards lower redshifts. 
Proper motions induced by the large 
gravitational potential of galaxy clusters
lead to the so-called  {\em fingers of god}
in the cone diagrams, i.e., 
elongated clumps with the major 
axis pointing toward the observer \cite[e.g.,][]{jac72}.
We find them preferentially in \class{2} 
cone diagrams, which we interpret as an
inclination for the red galaxies to be in 
clusters. \class{6} is formed by Seyfert
galaxies, and it seems to be more 
spreadout than the 
other classes (see Fig.~\ref{cone_diagram1}).
Finally, we find not distinct or sharp change of 
properties at redshift 0.25, i.e., at the 
divide used to split the classification
(see \S~\ref{final_imp}). Galaxies at
redshifts larger than this value
were not used to derive the classes.
The featureless transition at this special
redshift indicates no obvious systematic 
difference between the galaxies that 
define the classes, and the rest.

%
%%%%%
%
\section{Retrieving physical properties of individual
galaxies by interpolation}
\label{applications}

\begin{figure}
\includegraphics[width=0.45\textwidth]{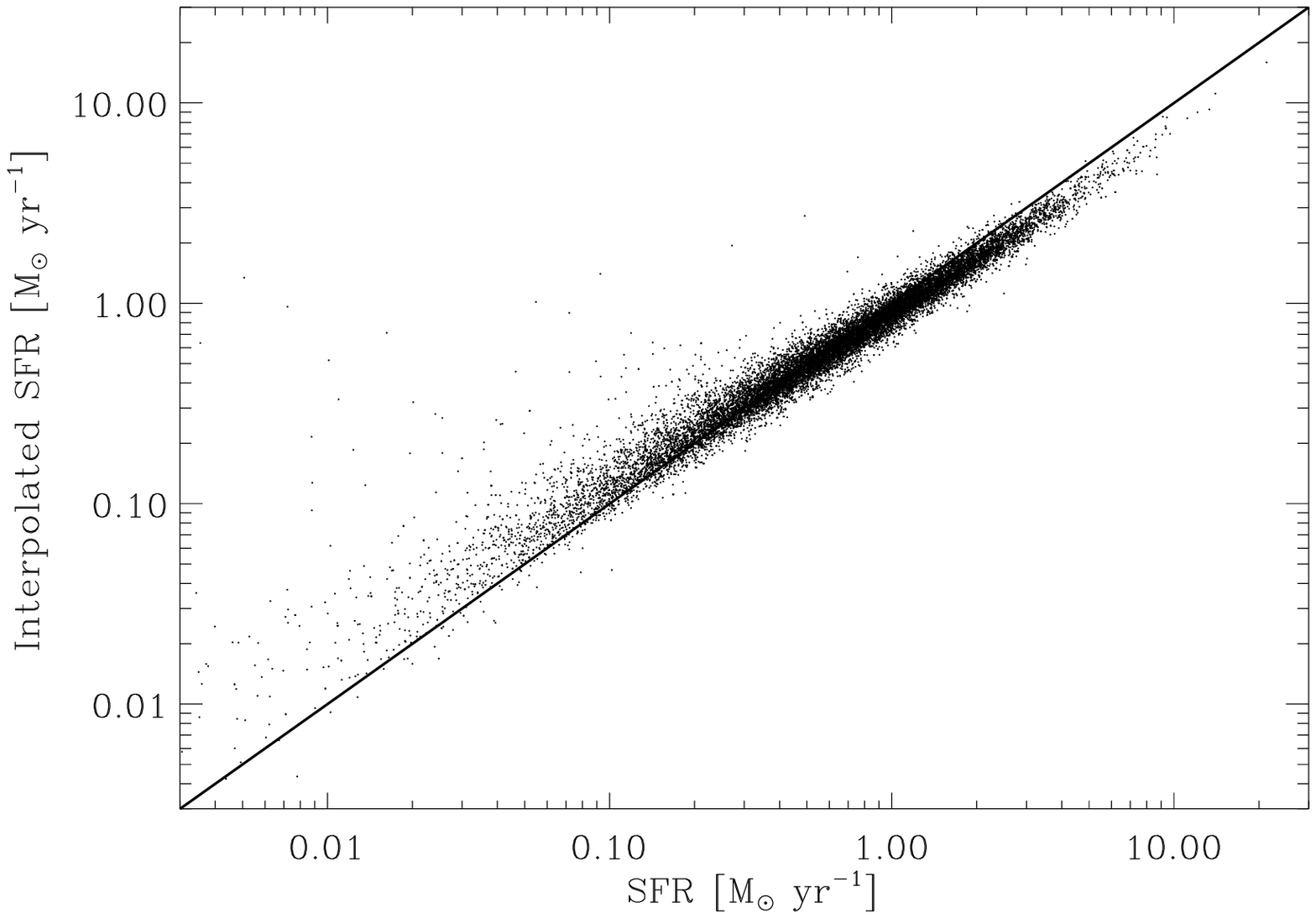}
\caption{
Scatter plot of Star Formation Rates (SFRs)
inferred by interpolation on the ASK classes 
(ordinates) versus SFR from the parameters
of the individual galaxies (abscissas).
The diagonal solid line indicates where abscissas and
ordinates coincide. Only a subset of randomly selected
starforming galaxies is represented (\class{$ \ge 7$}).
}
\label{fig_interpol}
\end{figure}

One can foresee several applications of the classification,
in particular,
it can be used to derive non-trivial physical parameters of 
individual galaxies by interpolation of the properties
of the classes. 
We want to measure the parameter $X$
for the galaxy ${\bf s}$. 
Assume that
the parameter $X$ varies systematically along the
ASK sequence, being $X_i$ in the $i$-th class.
Then, one can approximate $X$ for galaxy {\bf s} as,
\begin{equation}
X({\bf s})
\simeq {\sum_{i=0}^{27} Q_i({\bf s})\,X_i}\Big/
        {\sum_{i=0}^{27} Q_i({\bf s})},
\label{cutre_approx}
\end{equation}
where $Q_i({\bf s})$ represent the qualities
assigned to the galaxy as explained in \S~\ref{assigning},
except that we have shorten the notation so that,
\begin{equation}
Q_i({\bf s})=Q({\bf c_i},{\bf s},{\bf c_k}).
\end{equation}
In practice, the series~(\ref{cutre_approx})
can be truncated to consider only a few terms where the
quality is large enough. Regardless of 
the complications to estimate a given
parameter, by having
the classification and the value of the 
physical parameter in the ASK classes, 
equation~(\ref{cutre_approx})
trivially provides the parameter for all SDSS/DR7. 
We have used the Star Formation Rate (SFR)
to illustrate the procedure (Fig.~\ref{fig_interpol}).
The equivalent width of H$\alpha$ is a proxy
for Specific SFR (or SSFR) \citep[e.g.][]{ken98},
and it varies systematically along 
the ASK sequence (Table~\ref{tab2}).
Using the empirical relationship between H$\alpha$ flux 
and SFR as calibrated by \citet{ken98}, Fig.~\ref{fig_interpol}
shows a scatter plot with the SFRs
obtained directly and by interpolation based on
equation~(\ref{cutre_approx}). The figure considers
only starforming galaxies, i.e., 
\class{$ \ge 7$} according to \S~\ref{ask_vs_agn}. 
We truncate the series using only the three classes of 
highest qualities. Figure~\ref{fig_interpol} shows
that interpolated SFRs are correct within a factor
of two for SFRs varying four orders of magnitude. 
Unfortunately, the interpolation does not work in all
cases. We failed to estimate metallicities by 
interpolation. The oxygen metallicity
has a large dispersion within each ASK class and,
therefore, it does not meet the condition
of varying systematically along the ASK sequence.

The interpolation may be specially useful when dealing
with high redshift 
objects for which only noisy spectra are available.
Once each spectrum is  assigned to a particular class,
one can assign all the properties of the class
to the spectrum. In this sense, one can use the spectra of the
different classes as templates for 
redshift determinations \citep[e.g.,][]{lef05,lil07}. 
They represent a unique set able to reproduce all kinds
of  local galaxies. Unfortunately, the available
redshift range is limitted since the bluest 
wavelength of the templates is as red as 
3800\,\AA , and galaxies with redshift $ > 2.5$
will not overlap in any wavelength with 
the templates. However, one could overcome this problem 
classifying galaxies at various redshift
ranges in various steps, very much in the vein 
of the classification for galaxies with
redshift $> 0.25$ explained in \S~\ref{final_imp}.
After classification, the 
blue part of the average high redshift 
spectra can be used to provide blue parts for the 
templates. New templates with extended blue
wavelengths would be available, which can be used 
to extend the range even further by repeating the 
previous step. 
One can also use the classification to carry out 
relative measurements. For example, even if we ignore
the metallicity of a galaxy, one can infer whether
it is metal rich or poor with respect to the class mates 
using simple tools like the ratio between the 
fluxes of [NII]$\lambda$6583 and H$\alpha$
\citep[e.g.,][]{pet04}. All the systematic 
errors involved in such comparison would be 
greatly reduced within a class 
\citep[e.g.,][]{sta04,san09}.
Moreover, if the absolute metallicity of the class
template is known, these simple recipes 
for relative measurements yield absolute metallicities using
the class spectrum for reference.

%
%
%%%%%
\section{Discussion and conclusions}\label{conclusions}

We present an automatic unsupervised classification
of all the galaxies with spectra in the final SDSS data 
release (DR7). It uses the \kmeans\ algorithm, which separates 
the 930000 galaxies into 17 major classes containing 99\% 
of the galaxies, plus another 11 minor classes with the rest.
The algorithm guarantees that the galaxies in a class
have similar spectra, independently of 
their luminosities.
The algorithm does guarantees that
the classes represent true clusters in the
classification space, nor that all existing 
clusters are identified. 
Each ASK class\footnote{Acronym for Automatic Spectroscopic 
K-means-based class.} is characterized by an extra-low
noise spectrum resulting from averaging all the spectra in the 
class. These template
spectra vary smoothly and systematically among the classes,
labeled according to their $g-r$ color from ASK~0,
the reddest, to ASK~27, the bluest
(Fig.~\ref{classification} and Table~\ref{tab2}). 
The classes are well separated in the color sequence, 
with a class that collects most of the red sequence  
galaxies (ASK~2), a set of classes lying along the blue 
cloud (ASK 9 and larger), and a class that seems to
be characteristic of the green valley (ASK~5); 
see \S~\ref{colorscolors}. Usually the classes
of red galaxies do not present emission lines, however,
when they do, their excitation is characteristic
of AGN activity. In contrast, all the galaxies in the 
classes on the blue cloud seem to present emission lines,
but they are typical of star formation regions. The classes 
in between (i.e., in the green valley) show AGN activity 
(see \S~\ref{ask_vs_agn}). The ASK classification has 
been compared with the morphological Hubble type.
Although the number of galaxies involved in this 
comparison is rather limited, it clearly shows how the red 
classes tend to have early morphological types, 
whereas the blue classes are morphologically late types 
(Fig.~\ref{assign_kennicutt} and \S~\ref{ask_class_vs_morph}).
The relationship has a large intrinsic scatter, as 
previous studies also find (see \S~\ref{intro.intro}).
We have confronted the ASK classes with the PCA-based 
spectroscopic classification also existing for SDSS/DR7
(\S~\ref{pca_class}). The two of them are 
consistent in the sense that ASK classes have 
well defined PCA eigenvalues. However, the ASK classes
are finer. We note that the scatter between these two 
purely spectroscopic classifications is much smaller 
than the scatter in the relationship with  
Hubble type (\S~\ref{ask_class_vs_morph}). 
The distribution of classes
with redshifts is studied in \S~\ref{ask_vs_redshift},
and it reveals that the bluer classes contain galaxies
of lower redshift, indicating that they are made of 
galaxies less luminous (and so smaller) 
than the red classes.
The same preliminary analysis also 
suggests a trend for the red classes to be more
clustered than the blue classes.

All the above properties prove the consistency of 
the ASK classification. We have not found obvious 
contradictions between the physical properties of the 
classes, and the present understanding of galaxy properties.
However, one should not forget the limitations of the 
analysis. The classification is not unique. 
We know that the borders between classes are not 
well defined, and that the 
actual number of classes is somewhat arbitrary 
(\S~\ref{final_imp}; \S~\ref{test_qbcd}). 
The galaxy spectra seem to have a continuous 
distribution of properties, and we still ignore 
the reasons why \kmeans\ puts the borders between 
classes where it does. Moreover, the classification rely 
on a number of reasonable but otherwise subjective
hypotheses (e.g., the spectral bandpasses
entering into the classification, or the normalization
to the g-band; see \S~\ref{data_set} and Table~\ref{tab1}). 
Alternative hypotheses would render classifications 
differing from ASK in a way difficult to foretell.
All these caveats notwithstanding, 
the classes inferred by ASK have different spectra 
that reflect a systematic difference in the gas 
and stars present in the galaxies.  
Understanding the physical reasons causing
the systematic differences between spectra
would help understanding the classification 
itself.  
% However, the separation of galaxies in classes is only an 
% intermediate step. 
%  why there are 
% (only) these classes, and 
% One would like to know
% what are the underlying physical mechanisms that cause 
% the systematic differences between spectra.
% 
% We know that the borders between classes
% are not well defined, and that the actual number of 
% classes is somewhat arbitrary (\S~\ref{final_imp}; 
% \S~\ref{test_qbcd}). 
% 
The study of the
physical causes responsible for the observed diversity 
represents a major task that clearly goes beyond 
the scope of our introductory paper. However, a number of follow-up 
works dealing with the physical interpretation of classes are 
underway. 
As a result, some of the classes may need to
be joined or split.  For example, spectra of 
similar objects with different degrees of extinction 
may have ended up in different classes 
(remember that we do not correct for extinction 
to attain a purely empirical classification; 
\S~\ref{data_set}). 
Similarly, some of the classes may contain unidentified 
clusters.  Sub-divisions can be achieved 
by applying \kmeans\ to selected spectral windows
\citep[e.g., the region around H$\alpha$ may help us
separating emission line galaxies according 
to their metallicity;][]{pet04}. 
Deriving the star formation history of the 
classes is fundamental to understand whether 
the differences between spectra 
are tracing different star formation 
histories, AGN activities, merging histories, or something else. 
(Is the ASK classification revealing some sort
of evolutive sequence for galaxies?)
Fortunately, inversion codes able to constrain 
the star formation history are available
\citep[e.g.,][]{cid05,toj09}, and we plan to use them.
We are also carrying out a comparison between 
morphological types and spectroscopic types 
in a way that completes the introductory exercise in
\S~\ref{ask_class_vs_morph}. 
We try to understand what causes the scatter in the relationship between
morphology and spectroscopy. Is it the different
characteristic time of evolution of morphological
changes (on short timescales) and spectroscopic changes
(on long timescales)?
Is it the environment?
The morphological classification will be based
on the automatic procedure by \citet{hue08} using support 
vector machines, which will allow us to afford comparing 
morphology and spectral type for a 
sizeable fraction of the SDSS/DR7 spectroscopic 
catalog. 
Work to derive the 
luminosity function for the classes is 
pending, i.e., to characterize
the number density of galaxies of each luminosity and
class. It is needed to quantify the tendency 
for high ASK classes to contain dwarf galaxies, as
suggested in \S~\ref{ask_vs_redshift}.

In addition to understanding the physical mechanisms 
responsible for the diversity among spectra,
we foresee other applications of the ASK classification.
It provides a crude but fast way of estimating some
physical properties of a galaxy once its ASK
ascription is known (\S~\ref{applications}). 
The classification is also useful as target selection. 
For example, ASK~6 is formed by Seyfert galaxies. This 
class provides an ideal homogeneous sample of some 5000 
Seyferts with similar spectra for in-depth AGN studies 
\citep[e.g., extending to low
mass the relationship between supermassive black-hole
mass and bulge mass; see][and references therein]{fer06}.
The classification supplies classes of galaxies in the 
green valley (ASK~5). These targets allow us addressing 
the question of what characterizes a green valley galaxy,
and one can do it in a statistically significant way. 
Is the green valley a short period during the life of any galaxy, 
or does it represent a genuine class of galaxies separated
from the rest? 
The qualities assigned to each galaxy provide
a simple way to find unusual objects.
Low quality galaxies are outliers of the classification
and, therefore, abnormal objects that deserve
specific follow-up work.
The average spectra of the classes can be used as template
for redshift determinations.  They represent a unique  
set comprising all spectral types. 
This application of ASK requires  extending the template 
spectra to the UV,  but this upgrade can be done in successive 
steps as outlined in \S~\ref{applications}.

In order to facilitate these and possibly other
applications, we have made the ASK classification 
freely available though the ftpsite {\tt ftp://ask:galaxy@ftp.iac.es/}. 
We explain how it can be directly employed in SQL queries that use 
the CasJob facility of SDSS/DR7. We also
provide it as ASCII csv tables suitable for 
uses external to SDSS. In addition, the template spectra 
are included.

% \newpage
% 
% \begin{itemize}
% 
% %
% 
% \item Effects of aperture negligible \citep{yip04,zar95}.
% 
% 
% \item No further correction has been applied to the 
% data. We do not correct for extinction, seeing,
% galaxy size, aperture bias, etc. This apparent
% sloppiness actually results from 
% a deliberate attitude towards classification,  
% in the vein of the guidelines by \citet{san05} discussed
% in \S~\ref{intro}. If these corrections are 
% important and  the classification is working properly, 
% then the spectra of the same type of galaxies with and
% without an uncorrected bias should appear 
% in separate bins. It is then a matter of 
% a posteriori physical interpretation 
% to infer what causes the different bins,
% and eventually join them.
% 
% \end{itemize}
%%%%%%%

%\acknowledgement
\bigskip
\bigskip
\noindent{\bf Acknowledgments.}
We are indebted to J.~Betancort, 
I. G.~de la Rosa, and M.~Moles, that 
contributed with discusions on their 
area of expertise.
Thanks to an anonymous referee we include
the discussion on the degree of clustering of 
the classes at the end of \S~\ref{final_imp}.
The work has been partly funded by the Spanish
Ministries of science, technology and innovation, 
projects AYA~2007-67965-03-1, 
AYA~2007-67752-C03-01,
and CSD2006-00070.
    Funding for the Sloan Digital Sky Survey (SDSS) and SDSS-II has been provided 
    by the Alfred P. Sloan Foundation, the Participating Institutions, 
    the National Science Foundation, the U.S. Department of Energy, 
    the National Aeronautics and Space Administration, 
    the Japanese Monbukagakusho, and the Max Planck Society, 
    and the Higher Education Funding Council for England. 
    The SDSS is managed by the Astrophysical Research Consortium (ARC) 
    for the Participating Institutions. The Participating Institutions are the 
    American Museum of Natural History, Astrophysical Institute Potsdam, 
    University of Basel, University of Cambridge, Case Western Reserve University, 
    The University of Chicago, Drexel University, Fermilab, 
    the Institute for Advanced Study, the Japan Participation Group, 
    The Johns Hopkins University, the Joint Institute for Nuclear Astrophysics, 
    the Kavli Institute for Particle Astrophysics and Cosmology, 
    the Korean Scientist Group, the Chinese Academy of Sciences 
    (LAMOST), Los Alamos National Laboratory, the Max-Planck-Institute for 
    Astronomy (MPIA), the Max-Planck-Institute for Astrophysics (MPA), 
    New Mexico State University, Ohio State University, 
    University of Pittsburgh, University of Portsmouth, 
    Princeton University, the United States Naval Observatory, 
    and the University of Washington.

% 

% {\bf
% Copies to be send to Silk, this colleague of Alfonso that
% does the photo redshift in Alhambra, \dots.
% }
% 
% {\bf FTP site needs to be checked and completed}

\appendix

\section{Galaxies in a  cluster as a function 
        of the cluster center}\label{change_center}

The random initialization of \kmeans\ leads to 
small uncertainties in the properties of the clusters 
which,
however, produce a significant variation on the actual 
galaxies assigned to each cluster (\S~\ref{algorithm}, \S~\ref{final_imp}). 
This amplification can be understood by considering that 
clusters are defined by regions in a space of many dimensions.
A small uncertainty in the center of 
the cluster produces large (relative) variations of
the region of space that the cluster samples.
Below we compute this boost factor under the simplifying
assumption that the 
clusters are defined by hyper-spheres. Such simplification should not
affect the conclusion we draw since the scaling 
relationship between volume and 
area is a general property of the space, rather
then specific to a particular shape.

Assume that the galaxies belonging to a class are
those within a sphere of radius $R$ centered in the 
class center. Assume that the galaxies are uniformly
distributed around this center. Two different runs of 
the \kmeans\ clustering algorithm 
yield slightly different centers for the class,
separated by a distance $\Delta R$. 
How many galaxies 
will be shared by the two classifications?
Under the previous hypotheses, it is just the 
overlapping volume of two $n-$dimensional
hyper-spheres of radius $R$ when their centers are 
separated $\Delta R$. When $\Delta R /R\ll 1$, 
such volume is the volume of one of the 
original hyper-spheres minus the volume of a 
cylinder of height $\Delta R$ and base the 
corresponding $n-$dimensional hyper-disk 
(i.e., an hyper-sphere in $n-1$ dimensions).
Using the expression for the volume of a sphere in
$n$ dimensions, the number of common
galaxies, $N(\Delta R)$, normalized to the
number of galaxies in the class, $N(0)$,
turns out to be,
\begin{equation}
{{N(\Delta R)}\over{N(0)}}\simeq 
1-{{\Delta R}\over{R}}\sqrt{n/2\pi}.
\label{sphere}
\end{equation}
Equation~(\ref{sphere}) shows 
how a small relative error in the position 
of the cluster center gets amplified as 
$\sqrt{n/2\pi}$ when affecting the drop 
in the number of common galaxies. 
Since $n \gg 1$, the drop
is very large. For example,  if $n = 2000$, a 
minute $\Delta R/R\simeq 2$\% produces  
$N(\Delta R)/N(0)\simeq$ 65\%. Variations
induced by cluster
radius changes 
are even more dramatic. In this case
the boost factor scales
with $n$ rather than $\sqrt{n}$.

% \newcommand\aj{AJ}
% \newcommand\apj{ApJ}
% \newcommand\apjl{ApJ}
% \newcommand\apjs{ApJS}
% \newcommand\mnras{MNRAS}
% \newcommand\aapr{A\&ARev}
% \newcommand\araa{ARA\&A}
% \newcommand\aap{A\&A}
% \newcommand\nat{Nature}
% \newcommand\pasp{PASP}
% \bibliography{/home/jos/texto/papers/references/galax}
\bibliographystyle{aa}

\end{document}